\documentclass{aa}  

\usepackage{graphicx}
\usepackage{txfonts}
\usepackage[acronym,shortcuts]{glossaries} 
\usepackage[dvipsnames]{xcolor} 
\usepackage{booktabs}
\usepackage{newtxtext,newtxmath}
\usepackage{bm}
\usepackage{amsmath}	
\usepackage{multirow}
\glsdisablehyper 
\usepackage{ulem} 
\usepackage{xspace}
\usepackage[dvipsnames]{xcolor} 
\usepackage{caption}
\usepackage{hyperref}
\usepackage{orcidlink}
\usepackage{placeins}
\newcommand{\Rsun}{\ensuremath{\,\rm{R}_{\odot}}\xspace}
\newcommand{\Rj}{\ensuremath{\,\rm{R}_{J}}\xspace}

\newcommand{\kms}{\ensuremath{\,\rm{km}\,\rm{s}^{-1}}\xspace}
\newcommand{\gcc}{\ensuremath{\,\rm{g}\,\rm{cm}^{-3}}\xspace}
\newcommand{\Msun}{\ensuremath{\,\rm{M}_{\odot}}\xspace}
\newcommand{\Mj}{\ensuremath{\,\rm{M}_{J}}\xspace}
\newcommand{\Me}{\ensuremath{\,\rm{M}_{\oplus}}\xspace}

\newcommand{\Lsun}{\ensuremath{\,\rm{L}_{\odot}}\xspace}

\newcommand{\AU}{\ensuremath{\,\mathrm{AU}}\xspace}
\newcommand{\Myr}{\ensuremath{\,\mathrm{Myr}}\xspace}
\newcommand{\yr}{\ensuremath{\,\mathrm{yr}}\xspace}
\newcommand{\dy}{\ensuremath{\,\mathrm{d}}\xspace}
\newcommand{\hr}{\ensuremath{\,\mathrm{hr}}\xspace}

\newcommand{\Li}{\ensuremath{\mathrm{Li}}\xspace}

\newcommand{\ALi}{\ensuremath{\mathrm{A(Li)}}\xspace}
\newcommand{\Phantom}{{\scshape phantom}\xspace }

\newcommand{\MESA}{{\scshape mesa}\xspace }
\newcommand{\BV}{Brunt-V\"{a}is\"{a}l\"{a}\xspace}
\newcommand{\Mp}{\ensuremath{M_\mathrm{p}}\xspace}
\newcommand{\Rp}{\ensuremath{R_\mathrm{p}}\xspace}
\newcommand{\Mstar}{\ensuremath{M_\star}\xspace}
\newcommand{\Rstar}{\ensuremath{R_\star}\xspace}
\newcommand{\vKep}{\ensuremath{v_\mathrm{K}}\xspace}

\newacronym{BHL}{BHL}{Bondi-Hoyle-Lyttleton}
\newacronym{HJ}{HJ}{hot Jupiter}

\newcommand{\SPH}{SPH\space}

\begin{document} 

   \title{Hot Jupiter engulfment by an early red giant in 3D hydrodynamics}


   \author{Mike Y. M. Lau \inst{1,2,3,4}\fnmsep\thanks{\email{mike.lau@h-its.org}}\orcidlink{0000-0002-6592-2036} \and
           Matteo Cantiello \inst{4,5}\orcidlink{0000-0002-8171-8596} \and
           Adam S. Jermyn \inst{4}\orcidlink{0000-0001-5048-9973} \and \\
           Morgan MacLeod \inst{6}\orcidlink{0000-0002-1417-8024} \and
           Ilya Mandel \inst{3,2}\orcidlink{0000-0002-6134-8946} \and
           Daniel J. Price \inst{3}\orcidlink{0000-0002-4716-4235}}

   \institute{
      Heidelberger Institut f\"{u}r Theoretische Studien, Schloss-Wolfsbrunnenweg 35, 69118 Heidelberg, Germany \and
      OzGrav: The ARC Centre of Excellence for Gravitational Wave Discovery, Australia \and
      School of Physics and Astronomy, Monash University, Clayton, Victoria 3800, Australia \and
      Center for Computational Astrophysics, Flatiron Institute, 162 5th Avenue, New York, NY 10010, USA \and
      Department of Astrophysical Sciences, Princeton University, Princeton, NJ 08544, USA \and
      Center for Astrophysics | Harvard \& Smithsonian, 60 Garden Street, MS-16, Cambridge, MA 02138, USA
   }


  \abstract{
  Hot Jupiters are gas giant planets with orbital periods of a few days and are found in 0.1-1\% of Sun-like stars. They are expected to be engulfed during their host star's radial expansion on the red giant branch, which may account for observed rapidly rotating and chemically enriched giant stars. We performed 3D hydrodynamical simulations of hot Jupiter engulfment by a 1\Msun, 4\Rsun early red giant. Our `global' simulations simultaneously resolve the stellar envelope and planetary structure, modelling the hot Jupiter as a polytropic gas sphere. The hot Jupiter spirals in due to ram-pressure drag. A substantial fraction of its mass is continuously ablated in this process, although the mass-loss rate is resolution dependent. We estimate that this could enhance the surface lithium abundance by up to $\approx 0.1$ dex. The hot Jupiter is disrupted by a combination of ram pressure and tidal forces near the base of the convective envelope, with the deepest material penetrating to the radiative zone. The star experiences modest spin-up ($\sim 1\kms$), and engulfing a more massive companion may be required to produce a rapidly rotating giant. Drag heating near the surface and hydrogen recombination in the small amount of unbound ejecta recorded in the simulation could power an optical transient, although this needs to be confirmed by a calculation that has adequate resolution at the stellar surface.
  }

   \keywords{
      planet–star interactions -- planets and satellites: gaseous planets -- stars: low-mass -- hydrodynamics -- stars: chemically peculiar -- methods: numerical
   }

   \titlerunning{Hot Jupiter engulfment in 3D}
   \maketitle

\section{Introduction}
\label{sec:intro}
Transit, radial velocity, direct imaging, and microlensing searches have uncovered more than 5,000 confirmed exoplanets over the past two decades, providing valuable insight into their demographics. A substantial fraction of exoplanets are found to have short orbital periods. In particular, \acp{HJ} are gas giant planets with orbital periods of a few days, and were the first exoplanets discovered around main sequence stars \citep{Mayor&Queloz+95,Dawson&Johnson+18,Fortney+21}. Both transit-based \citep{Gould+06,Bayliss+11,Howard+12a,Fressin+13} and radial-velocity-based surveys \citep{Marcy+05,Cumming+08,Mayor+11,Wright+12} find that a few times 0.1 to $\approx1\%$ of Sun-like stars host \acp{HJ}.

While the origin of their short orbital periods is interesting by itself, the close proximity of \acp{HJ} to their host stars also means they should be engulfed by the radial growth of their host stars, up to an \AU on the red giant branch. This is estimated to occur at a rate of 0.1--1 yr$^{-1}$ in the Galaxy, based on the $\sim 1\%$ occurrence rate of \acp{HJ} \citep{Metzger+12,MacLeod+18}. Processes that shrink a planet's orbit cause the planet to become engulfed beneath their present-day separations. Close-in companions ($\lesssim 0.1\AU$) can be drawn in by tidal dissipation, particularly during post-main-sequence expansion \citep{Rasio+96,Villaver&Livio07,Villaver&Livio09,Jackson+09,Miller+09,Kunimoto+11,Barker20,Lazovik21}. Farther out companions can be brought in via high-eccentricity migration, where the orbit is first driven to be highly eccentric, followed by tidal circularisation at periastron. Mechanisms for increasing the orbital eccentricity include planet-planet scattering \citep{Rasio&Ford96,Weidenschilling&Marzari96,Juric&Tremaine08,Nagasawa&Bessho08} and (eccentric) Kozai-Lidov oscillations induced by a planetary or stellar companion \citep{Eggleton&Kiseleva-Eggleton01,Wu&Murray03,Fabrycky&Tremaine07,Wu+07,Naoz+11,Naoz+12,Stephan+18}.

Planetary engulfment could produce a number of observable signatures. The engulfed planet's orbital decay may power a luminous transient \citep{Retter&Marom03,Metzger+12,MacLeod+18,Yarza+22,Gurevich+22}. Partial or complete dissolution of the planet in a giant star's convective envelope could chemically enrich the stellar surface \citep{Alexander67,Israelian02,Aguilera-Gomez+16,Casey+19,Soares-Furtado+21}, possibly explaining the existence of some lithium-enriched giant stars \citep[e.g.][]{Monroe+13,Carlos+16,Melendez+19,Nagar+19}. Observations and canonical stellar models find that surface Li is generally  depleted when ascending the red giant branch. This is due to the convective envelope's inward growth during the first dredge-up, which mixes Li-depleted material to the stellar surface. However, $\sim1\%$ of giant stars are Li rich (\ALi > 1.5\footnote{The relative Li abundance is defined as $\mathrm{A}(\mathrm{Li}) = 12 + \log_{10}[N(\mathrm{Li})/N(\mathrm{H})]$, where $N(\mathrm{E})$ is the number of atoms of the element with symbol E.}), and Li-super-rich (\ALi > 2.7) giant stars, which have super-meteoritic Li abundances, have also been observed. The angular momentum deposited during planetary engulfment could induce rotational mixing that reaches the H-burning shell, enriching the surface with Li via the Cameron-Fowler process \citep{Cameron&Fowler71}. The connection between planetary engulfment and these observations is supported by a possible correlation of Li enrichment with rotational velocity \citep{Carlberg+13} and planet occurrence \citep{Adamow+18}.

Understanding the ability for planetary engulfment to produce these effects requires determining the amount of mass lost by the planet in the convective envelope, the dynamical response of the star to the planet's spiral-in, and the distribution of angular momentum following engulfment. These effects are well suited to being studied with hydrodynamical simulations. However, the large dynamic range spanned by the engulfment process makes it a challenging computational problem, and so various approximations have been made in past simulations. \cite{Sandquist+98,Sandquist+02} focused on the local gas flow around engulfed planets, reporting significant dissolution within the stellar envelope. On the other hand, \cite{Staff+16} performed global simulations of a 10\Mj planet engulfed by a 3.5\Msun red giant and by a 3.05\Msun thermally pulsating asymptotic giant. They resolved the global envelope structure but approximated the planet as a point mass. Recently, \cite{Yarza+22} simulated the local, steady-state gas flow around an engulfed planet across a parameter space mapped out by analogy with the wind tunnel formalism developed for studying common-envelope interactions \citep{MacLeod&Ramirez-Ruiz15,MacLeod+2017}. A key difference is that the perturber's size is a significant fraction of its Bondi radius, and so it was modelled as a rigid sphere with a reflective boundary. \cite{Yarza+22} predict that planetary engulfment could increase the host star's luminosity by several orders of magnitude over $1-10^3\yr$, depending on the star's evolutionary stage and the planet mass.

In this study, we combined various aspects of previous approaches, and performed a 3D global hydrodynamical simulation of the engulfment of a 1\Mj \ac{HJ} by a 1\Msun early-stage red giant. We modelled the \ac{HJ} as a gas sphere in order to self-consistently model changes to the planet mass and structure and to include the effects of ram-pressure drag or hydrodynamical drag, the dominant form of drag in our scenario (see Section \ref{subsec:energy_mech}). \cite{Abia+20} applied a similar approach to modelling the collision between a brown dwarf and a 1\Msun main-sequence star.

This paper is structured as follows. In Sections \ref{subsec:regimes}, we outline different regimes of star-planet interaction to provide context for our chosen system parameters. In Section \ref{sec:methods}, we describe the system we simulated and the simulation setup. In Section \ref{sec:results}, we present our results, including the different phases of planetary engulfment (Section \ref{subsec:phases}), the evolution of the planet velocity (\ref{subsec:vel}), planet mass ablation (\ref{subsec:ablation}), and the disruption and penetration depth of planetary material (\ref{subsec:disruption}). In Section \ref{sec:discussion}, we discuss possible observational signatures associated with \ac{HJ} engulfment, including surface chemical enrichment (\ref{subsec:enrichment}), induced stellar rotation (\ref{subsec:spinup}), and luminous transients (\ref{subsec:transients}). We summarise our findings in Section \ref{sec:conclusion}.

\section{Star-planet interaction regimes}
\label{subsec:regimes}
The regimes of star-planet interaction may be broadly characterised by the mass ratio, $\Mp/\Mstar$, and radius ratio, $\Rp/\Rstar$, of the planet and host star. Figure \ref{fig:regimes} shows the different interaction regimes in $\Mp/\Mstar$--$\Rp/\Rstar$ space. The blue markers indicate engulfment events that would be experienced by confirmed exoplanets in the NASA Exoplanet Archive \citep{Akeson+13} when their host stars expand to their present-day orbital separations (i.e. taking $R_\star\rightarrow a$). This represents an upper limit to the stellar radius at the moment of engulfment, in the limit of inefficient orbital decay due to tides. The plotted sample comprises 1,491 exoplanets with measured \Rp, \Mp, \Mstar, and $a$. The \acp{HJ} can be seen to cluster at $\Mp/\Mstar\sim10^{-3}$ and $\Rp/a\sim10^{-2}$, whereas the hot Neptunes populate $\Mp/\Mstar\sim10^{-5}-10^{-4}$ and $\Rp/a\sim10^{-3}$.

\subsection{Energy dissipation mechanism}
\label{subsec:energy_mech}
An engulfed planet spirals in as orbital energy is dissipated into the stellar envelope by drag forces. The spiral-in timescale may be estimated as the ratio of the orbital energy to the drag luminosity, $|E_\mathrm{orb}|/\dot{E}_\mathrm{drag}$. The drag luminosity generally scales as $\dot{E}_\mathrm{drag} \sim \rho R_\mathrm{eff}^2 v^3$, where $\rho$ is the local density of the background stellar gas, $v$ is the planet velocity relative to this background gas, and $R_\mathrm{eff}^2$ is the cross-section for interacting with the background medium. This cross-section depends on the dominant form of drag, which can be gravitational drag \citep{Chandrasekhar43,Ostriker1999} or turbulent hydrodynamical/aerodynamic drag. Gravitational drag originates from gravitationally focusing upstream material into a high-density wake that trails the planet. The interaction radius is the planet's \ac{BHL} radius \citep{Hoyle-Lyttleton1939,Bondi1952}, which scales as $R_\mathrm{eff} = R_\mathrm{BHL} \sim G\Mp/v^2$ when the flow is supersonic. On the other hand, hydrodynamical drag arises from the ram pressure of the upstream flow, and so the interaction radius is approximately the planet's geometric radius, $R_\mathrm{eff} \approx \Rp$.

We can write down the approximate scaling for the dissipation timescale by taking $\rho\sim\Mstar/\Rstar^3$ and assuming a Keplerian orbit such that $E_\mathrm{orb} \sim -G\Mstar\Mp /\Rstar$ and $v \sim (G\Mstar/\Rstar)^{1/2}$. Then, the timescale for energy dissipation by hydrodynamical drag is
\begin{align}
    \frac{|E_\mathrm{orb}|}{\dot{E}_\mathrm{hydro}} \sim \biggl( \frac{\Mp}{\Mstar} \biggr) \biggl( \frac{\Rp}{\Rstar} \biggr)^{-2} t_\mathrm{dyn},
    \label{eq:thydro}
\end{align}
where $t_\mathrm{dyn} = [\Rstar^3/(G\Mstar)]^{1/2}$ is the star's surface free-fall time. Taking instead $R_\mathrm{eff} = R_\mathrm{BHL} \sim G\Mp/v^2$, the timescale for energy dissipation by gravitational drag is
\begin{align}
    \frac{|E_\mathrm{orb}|}{\dot{E}_\mathrm{BHL}} \sim \biggl( \frac{\Mp}{\Mstar} \biggr)^{-1} t_\mathrm{dyn}.
    \label{eq:tBHL}
\end{align}
The boundary between the two regimes is given by $\Rp = R_\mathrm{BHL}$, that is, when $\Rp/\Rstar = \Mp/\Mstar$, which is plotted as the solid line in Figure \ref{fig:regimes}. Most exoplanets, when engulfed by their host stars at their present-day separations, are located in the regime dominated by hydrodynamical drag. Gravitational drag is only relevant for the most massive \acp{HJ} and for brown dwarfs, or if engulfed by a large, evolved star \citep{Staff+16}. For a regular Jovian planet ($\Mp\sim10^{-3}\Msun$, $\Rp\sim0.1\Rsun$), gravitational drag is only important if it is engulfed at $\gtrsim 100\Rsun$.

In the gravitational drag regime, equation (\ref{eq:tBHL}) implies that the orbital separation shrinks over many orbits since $\Mp\ll\Mstar$. This is not necessarily true in the hydrodynamical regime, where equation (\ref{eq:thydro}) implies the inspiral can take place over several orbits when $\Mp/\Mstar\lesssim(\Rp/\Rstar)^2$. In the more extreme case where $\Mp/\Mstar\ll(\Rp/\Rstar)^2$, rather than experiencing a complete inspiral, the drag rapidly dampens the azimuthal orbital velocity and the planet transitions to nearly radial sinking, still mediated by drag.

\subsection{Planet Roche-lobe overflow}
\begin{figure}
    \includegraphics[width=\linewidth]{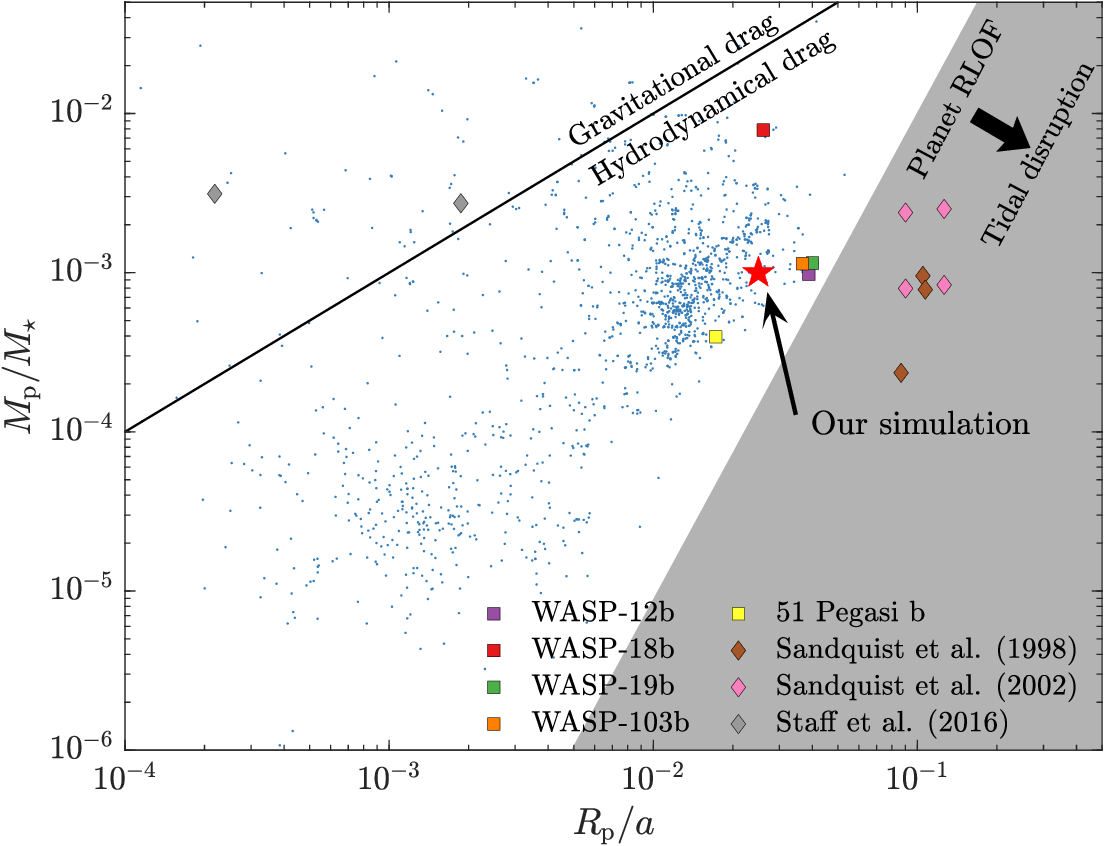}
    \caption{Star-planet interaction regimes in mass and radius ratio space. Blue points correspond to observed exoplanets in the NASA Exoplanet Archive \citep{Akeson+13}, assuming they will be engulfed when their host stars grow to their current separations ($\Rstar\rightarrow a$). Square markers highlight specific hot Jupiters while the diamond markers indicate planetary engulfment scenarios explored in past simulations. The solid line is the approximate boundary that determines whether gravitational drag or ram-pressure/hydrodynamical drag dominates in the engulfment process. The shaded area represents an unpopulated region where a planet would exceed its Roche lobe and engage in mass transfer, thus avoiding direct impact with the stellar surface. Planets that are dynamically scattered deep within this region would experience tidal disruption.}
    \label{fig:regimes}
\end{figure}

The shaded region in Figure \ref{fig:regimes} is unpopulated, as exoplanets in this region would exceed their Roche lobes. A planet that is driven inwards into this region via tidal dissipation would therefore avoid direct impact with the stellar envelope \citep{Metzger+12}. Instead, they would either gradually lose their gaseous envelopes through stable mass transfer \citep{Valsecchi+14,Valsecchi+15,Jackson+16,Ginzburg+17} or plunge towards the host star in dynamically unstable mass transfer \citep{Jia&Spruit17}.

We calculate the boundary of the `Roche-lobe overflow' regime by equating the planet's Roche limit with its present-day separation, $a_\mathrm{RLOF}(\Mp/\Mstar) = a$, using the \cite{Eggleton1983} approximation to evaluate the Roche radius as a function of mass ratio. For example, a 1\Mj and 1\Rj planet orbiting a 1\Msun star has $a_\mathrm{RLOF} = 0.01\AU$ ($P_\mathrm{orb}=0.4$ d). So solar-mass host stars that have evolved beyond the end of the subgiant branch, with radii $\Rstar \gtrsim 2\Rsun$, may directly engulf \acp{HJ} at their present separations. Engulfment at an earlier evolutionary phase requires the \ac{HJ} to migrate inwards towards the Roche limit \citep{Metzger+12} or to inflate due to tides and stellar irradiation \citep[e.g.][]{Miller+09,Komacek+20}. 

A number of \acp{HJ} have been observed to lie near this boundary, as highlighted by some of the square markers in Figure \ref{fig:regimes}. For example, WASP-12b (purple square), a 1.47\Mj and 1.9\Rj \ac{HJ} on a 1.09-day orbit \citep{Hebb+09}, is at 80\% of its Roche limit. There have been indications that its tenuous, optically-thin exosphere exceeds its Roche lobe and is being gradually transferred onto a debris disk around its host \citep{Li+10,Fossati+10}.

\subsection{Tidal disruption}
Planets that enter even smaller radii (inside the shaded region), whether induced by planet-planet scattering or secular chaos \citep{Mardling95}, or as a result of dynamically unstable mass transfer, could be partially or completely destroyed by tides over multiple periastron encounters, thereby also averting the direct engulfment scenario \citep{Faber+05,Guillochon+11,Liu+13,Guidarelli+19,Guidarelli+22}. This occurs near the tidal disruption radius, $r_\mathrm{td} = (\Mstar/\Mp)^{1/3}\Rp$. For a 1\Mj and 1\Rj planet around a 1\Msun star, $r_\mathrm{td} = 1\Rsun$. We note that $r_\mathrm{td}$ has the same scaling as the Roche limit, with a prefactor that depends on the planetary structure and the extent of disruption. Hydrodynamical simulations suggest that disruption could occur out to $2.7r_\mathrm{td}$ \citep{Guillochon+11}.

\section{Methods} \label{sec:methods}

\subsection{System parameters}
\label{subsec:parameters}
Based on the considerations in Section \ref{subsec:regimes}, we choose to simulate the engulfment of a 1\Mj, 1\Rj \ac{HJ} by a 1\Msun star\footnote{More precisely, the simulated \ac{HJ} has mass of 1.05\Mj and radius of 0.97\Rj. These values are chosen so that the planet-star mass ratio ($1.00\times10^{-3}$) and radius ratio ($2.50\times10^{-2}$) are simple fractions. This difference has little practical importance, and so we simply state the simulated \ac{HJ}'s initial mass and radius to a single significant figure throughout this paper.} that has evolved to a radius of 4\Rsun (0.019\AU), whereupon it is ascending the red giant branch, before the luminosity bump (Figure \ref{fig:HR}). The red star in Figure \ref{fig:regimes} shows where our chosen parameters lie in $\Mp/\Mstar$--$\Rp/\Rstar$ space. 

The chosen planet mass and radius are typical among the \ac{HJ} population. However, the 0.019\AU orbital separation (implying $P_\mathrm{orb} = 0.9\dy$) is 2--3 times smaller than the population median (see Figure \ref{fig:regimes}). Nonetheless, this is an interesting and relevant choice for several reasons. Firstly, previous hydrodynamical simulations have either focused on engulfment on notably smaller or larger separations. \cite{Staff+16} used red giant and asymptotic giant branch stars (grey diamond markers in Figure \ref{fig:regimes}), which are qualitatively different to our scenario as gravitational drag drives the spiral-in in their case. Moreover, their large luminosities and deep convective envelopes make it challenging to produce detectable luminous transients and statistically significant chemical enrichment signatures \citep[][and see also Section \ref{sec:discussion}]{MacLeod+18,Soares-Furtado+21}. On the other hand, the simulations by \cite{Sandquist+98,Sandquist+02} used main-sequence stars, which imply engulfment at even smaller separations. Figure \ref{fig:regimes} shows that these simulations (brown and pink diamond markers) are located deep within the shaded region. That is, the planet they studied would have experienced intense tidal forces, if not complete disruption, prior to entering the stellar envelope. Thus, we choose simulation parameters that lie safely outside the `Roche-lobe overflow' region.

Our chosen parameters match the handful of so-called ultra-short period \acp{HJ} that have been observed. For example, WASP-18b \citep{Hellier+09} and WASP-103b \citep{Southworth+15}, which are labelled in Figure \ref{fig:regimes}, have similar semi-major axes to our simulated system, measuring 0.02026 and 0.01987\AU, respectively. Our simulations could therefore reflect their eventual engulfment if they maintain their present-day separations. Alternatively, a \ac{HJ} could also be tidally captured from a wider initial orbit and engulfed \citep{Villaver&Livio09,Privitera+16a}. WASP-12b is to date the only \ac{HJ} with a measured orbital decay rate at $29.81\pm0.94~\mathrm{ms}~\mathrm{yr}^{-1}$, and decay timescale of $P_\mathrm{orb}/\dot{P}_\mathrm{orb}=3.16\pm0.10\Myr$ \citep{Wong+22}.

\subsection{Setup}
We performed 3D hydrodynamical simulations of planetary engulfment using the smoothed particle hydrodynamics \citep[SPH;][]{Monaghan92,Price12} code \Phantom \citep[v2022.0.1,][]{Price+18}. We resolve the star with $N_\star = 10^7$ \SPH particles (excluding the H-burning shell and the region interior to it, see Section \ref{subsec:star}) and the planet with 12,300 \SPH particles. The particles have equal masses and smoothing lengths proportional to $\rho^{-1/3}$, ensuring a fixed average number of particle neighbours. To check for convergence, we additionally performed a series of simulations at different resolutions ($N_\star = 10^6$, $3\times10^6$, $10^7$, and $3\times10^7$ \SPH particles), but with smaller initial separations to compensate for the increased computational cost of using higher resolutions than our default.

For all simulations, we use an ideal gas equation of state with an adiabatic index of 5/3. We do not include radiation transport, which is only significant on timescales much longer than our simulations (see Section \ref{subsec:star}). We use \Phantom's default shock viscosity switch based on \cite{Cullen+Dehnen10} and default artificial thermal conductivity parameter ($\alpha_u=1$). To alleviate the computational cost of simulating a large dynamic range, we allow \SPH particles to evolve on their individual time-steps, unlike our previous simulations that used global time-stepping \citep{Lau+22a,Lau+22b}. Our simulation conserves energy to within 1.0\% and angular momentum to within 0.9\%.

\subsection{Hot Jupiter model}
We model the \ac{HJ} as an $n=3/2$ polytropic gas sphere \citep{Hubbard84,Stevenson91}. Modelling the planet as a gas sphere rather than a point particle \citep[as in][]{Staff+16} or a rigid sphere \citep[as in][]{Yarza+22} is required in the hydrodynamical drag regime and also allows planetary ablation and tidal disruption to take place. Jovian planets have a heavy-element core with a supersolar-enriched metallic hydrogen outer layer, and a molecular hydrogen envelope enriched with heavy elements. Our polytropic model neglects this core-envelope structure, but is an adequate approximation since the core contains only a few percent of the total mass (few tens of \Me).

\subsection{Stellar model}
\label{subsec:star}
\begin{figure}
    \includegraphics[width=\linewidth]{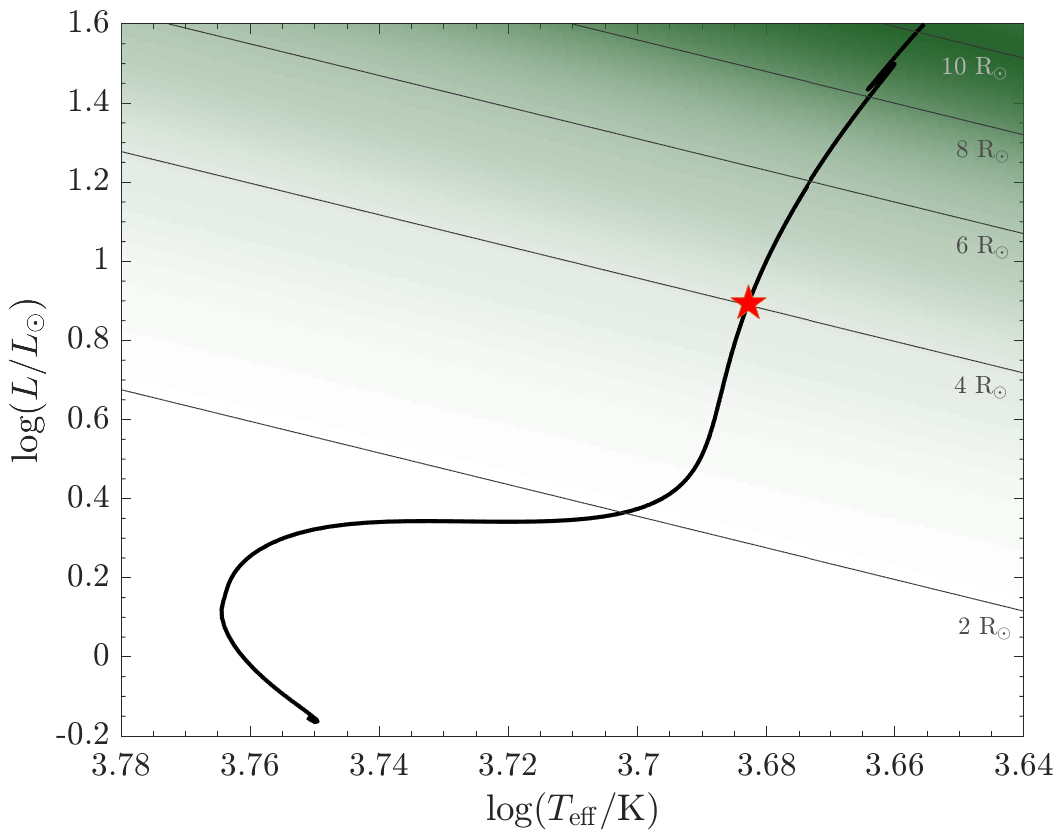}
    \caption{Evolutionary track of our 1\Msun stellar model on the H-R diagram, as calculated by \MESA. The red star shows the location of the 4\Rsun red giant model used in our simulation. The grey lines are constant radius contours. The intensity of green shading is proportional to the density of orbital separations in the NASA Exoplanet Archive.}
    \label{fig:HR}
\end{figure}

To obtain a model for the host star, we used the 1D stellar evolution code \MESA \citep{MESA1,MESA2,MESA3,MESA4,MESA5} to evolve a 1\Msun star to the red giant phase until reaching 4\Rsun in radius\footnote{The \MESA model and the inlists used to produce it are available at \url{https://dx.doi.org/10.26180/21342090}. We assume solar metallicity, $Z = 0.0142$, and step overshooting by 0.11 times the pressure scale height (\texttt{overshoot\_f = 0.11}), using the convective mixing diffusion coefficient at 0.01 times the pressure scale height interior to the step overshoot boundary (\texttt{overshoot\_f0 = 0.01}).}. Figure \ref{fig:HR} shows the evolution of our 1\Msun model on the H-R diagram, with the red star marking the selected model. This model has a luminosity of 7.77\Lsun and effective temperature of 4,820 K.

\begin{figure}
    \includegraphics[width=\linewidth]{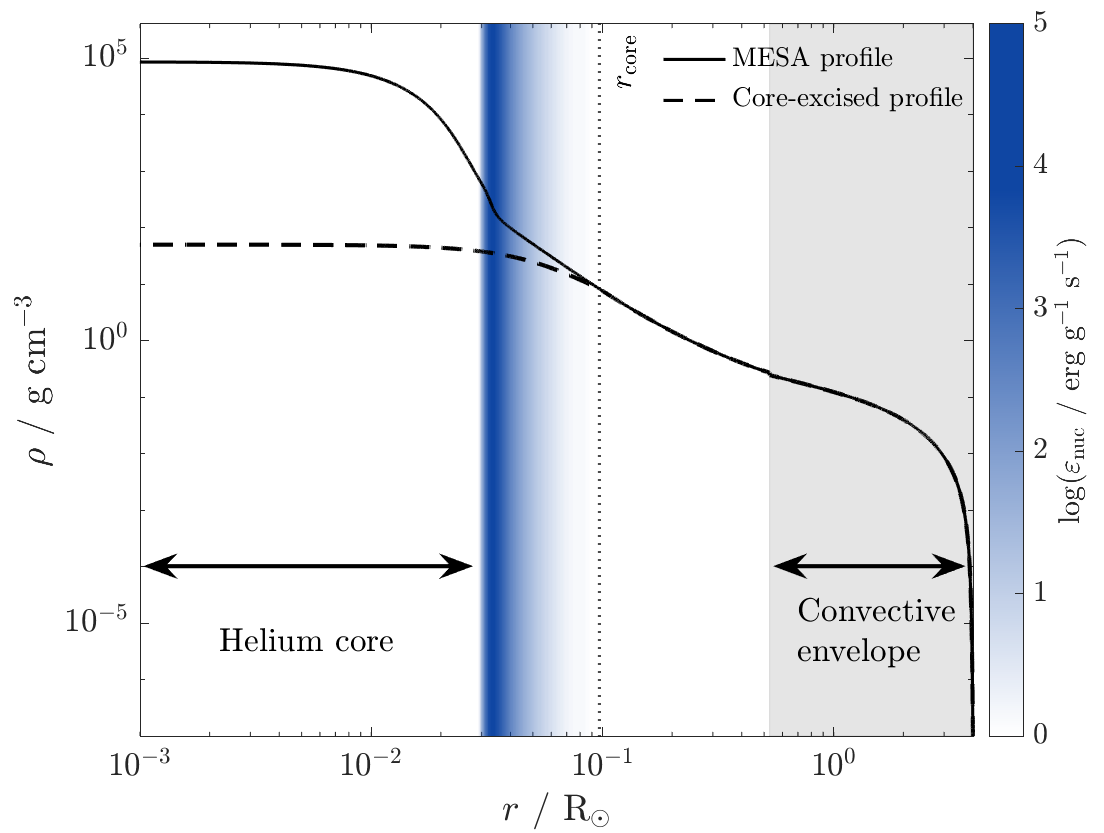}
    \caption{Density profile of the red giant host star as obtained from \MESA (solid line) and after excising the high-density `core' (dashed line). The two profiles are identical above $r_\mathrm{core} = 0.097\Rsun$ (dotted line). We also show the extent of the convective envelope (grey region) and the He core. The background is shaded blue according to the specific nuclear burning luminosity, $\varepsilon_\mathrm{nuc}$, showing the location of the H-burning shell.}
    \label{fig:profile}
\end{figure}

Figure \ref{fig:profile} shows its density profile (solid line). It consists of a 0.18\Msun inert He core that is $3.3\times10^{-2}\Rsun$ in size, and a 0.75\Msun and 3.5\Rsun deep convective envelope (shaded in grey), separated by a 0.50\Rsun radiative zone. The plot is shaded blue according to the specific nuclear burning luminosity, $\varepsilon_\mathrm{nuc}$, showing the location of the H-burning shell. Within the envelope, the convective turnover time ranges from $\sim 10$ days near the surface to $\sim 100$ days near the base of the convective zone. This is much longer than the duration we simulate (76\hr) and allows us to neglect the envelope's convective motion.

The stellar density spans 12 orders of magnitude across the star, which presents a severe dynamic range problem. As with most global hydrodynamical simulations involving giant stars, we replaced the dense stellar core, defined as the region beneath some radius $r_\mathrm{core}$, with a point mass that interacts via a softened gravitational potential. We used a 0.187\Msun point mass and core radius of $r_\mathrm{core}=0.097$\Rsun (dotted line in Figure \ref{fig:profile}), which falls between the H-burning shell and the base of the convective zone. The density profile beneath $r_\mathrm{core}$ must be replaced with a distribution that is in hydrostatic equilibrium with the softened point mass potential. We solved for this profile, chosen to have constant entropy, according to Section 3.1 of \cite{Gonzalez-Bolivar+22}\footnote{This procedure is automated as a setup option for stars in \Phantom.}. The dashed line in Figure \ref{fig:profile} shows this modified profile. As can be seen, our point mass replacement lowers the maximum stellar density that has to be resolved by three orders of magnitude.

For both the star and the planet, we used the technique developed by \cite{Lau+22a} (Appendix C) to map a 1D density profile into a 3D distribution of \SPH particles that exhibits a high degree of hydrostatic balance and accurately reproduces the intended density profile. We have verified that the host star, when simulated by itself, maintains its radius to within 0.1\% and its central density to within 1.2\% over the duration of our simulation (21 times the surface free-fall time). In Appendix \ref{app:planet-stability}, we show more explicitly that the planet also maintains its original structure.

\subsection{Initial orbit}
We assume that the initial star-planet orbit is circular and Keplerian. While many \acp{HJ} in the 3--10\dy period range are observed to have moderate eccentricities ($0.2<e<0.6$), consistent with dynamical origins, most with periods $\lesssim3\dy$ are consistent with being circular \citep{Dawson&Johnson+18}, as tidal circularisation becomes very efficient at these small separations \citep{Dobbs-Dixon+04}. We also assume that both the host star and the \ac{HJ} are initially non-rotating. The non-rotating assumption is usually made for expanded donors on the giant branch, which are expected to have surface rotational velocities that are a few \kms \citep[e.g.][]{Carlberg+11}. On the other hand, short-period \acp{HJ} are expected to have tidally synchronised with their orbits, but the small amount of angular momentum this would contribute ($\lesssim 0.1\%$ of the orbit) can be safely neglected. The initial separation was chosen such that the star and planet are just touching: $a(t=0) = \Rstar + \Rp = 4.1\Rsun$, with the exception of test simulations in our resolution study (Appendix \ref{app:resolution}), which begin with the planet fully immersed in the stellar envelope, $a(t=0) = 3.8\Rsun$.

\section{Results} \label{sec:results}

\subsection{Stages of planetary engulfment}
\label{subsec:phases}
\begin{figure*}
    \centering
    \includegraphics[width=0.98\linewidth]{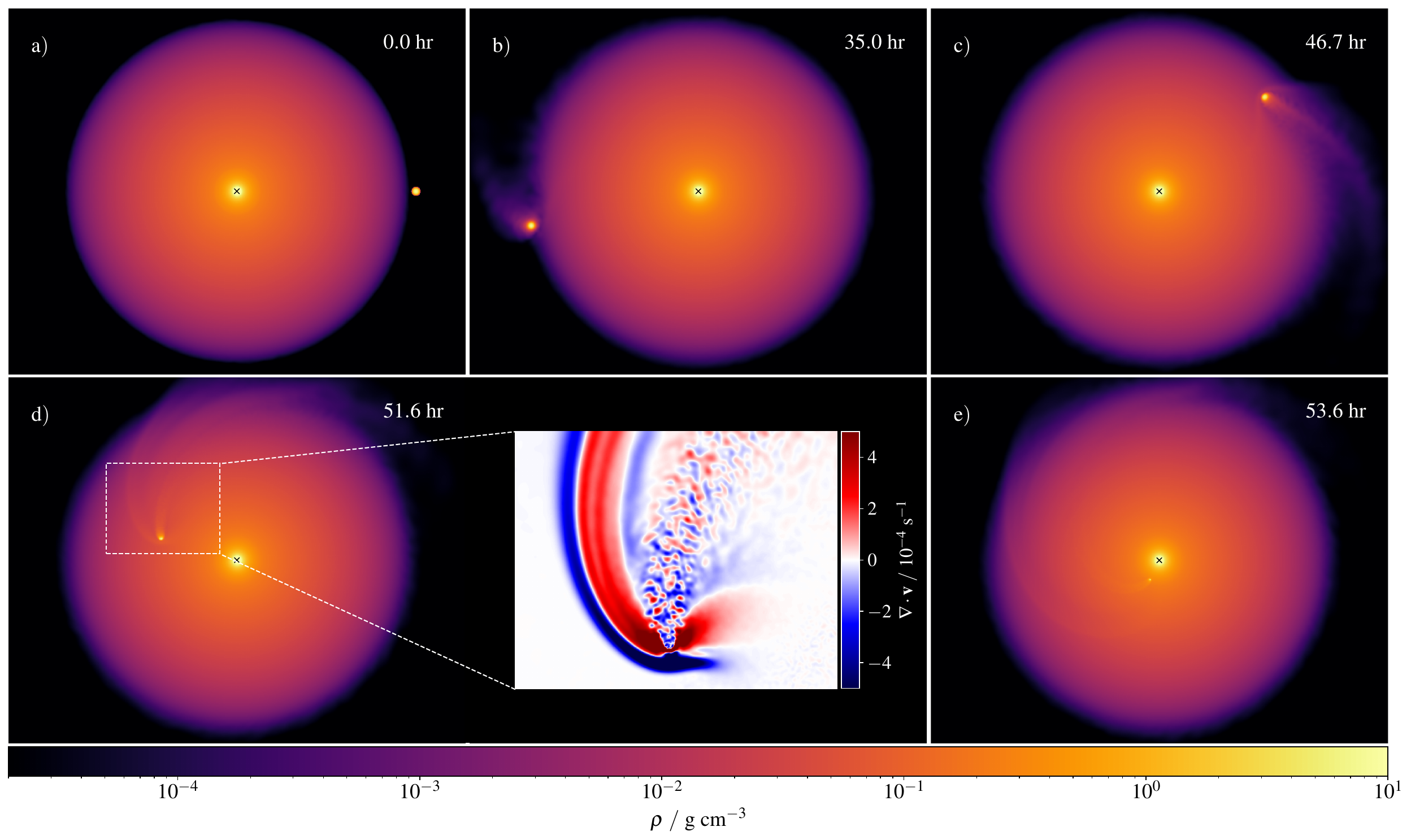}
    \caption{Density slices ($8.4\Rsun\times11.6\Rsun$) in the orbital plane. a): Start of the simulation where the hot Jupiter grazes the stellar surface ($P_\mathrm{orb}=23.1\hr$). b)-c): The hot Jupiter is fully engulfed and spirals in due to drag. d)-e): Radial sinking and disruption of the hot Jupiter. The inset in d) plots the velocity divergence, which shows the bow shock (dark blue) and downstream expansion region (red). The black cross marks the location of the sink particle used to replace the stellar core. This Figure was created with the \textsc{Sarracen} package \citep{Harris+Tricco23}. Videos of our simulations are available at \protect\url{https://themikelau.github.io/planet_engulfment.html}.}
    \label{fig:rho_xy}
    \includegraphics[width=0.98\linewidth]{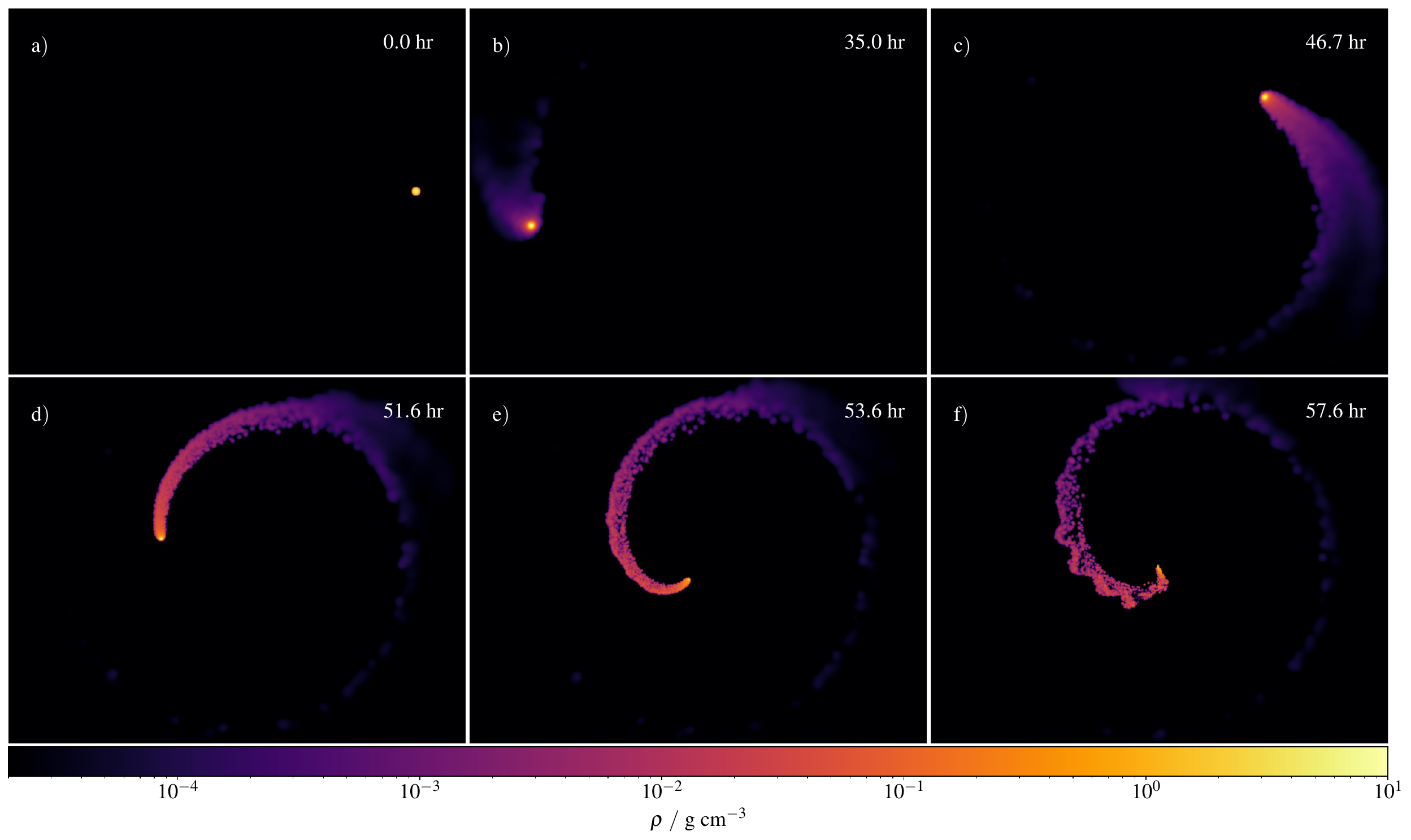}
    \caption{Similar to Figure \ref{fig:rho_xy}, but showing only the planetary material and including an additional panel f) near the end of the simulation. The observed granularity is an artefact of rendering a restricted number of \SPH particles.}
    \label{fig:rho_planet}
\end{figure*}

We present our simulation in Figures \ref{fig:rho_xy} and \ref{fig:rho_planet}, which show density slices in the orbital plane at different points in time. In Figure \ref{fig:rho_planet}, only \SPH particles that initially constituted the \ac{HJ} (`planet particles') are rendered.

\begin{figure}
    \includegraphics[width=\linewidth]{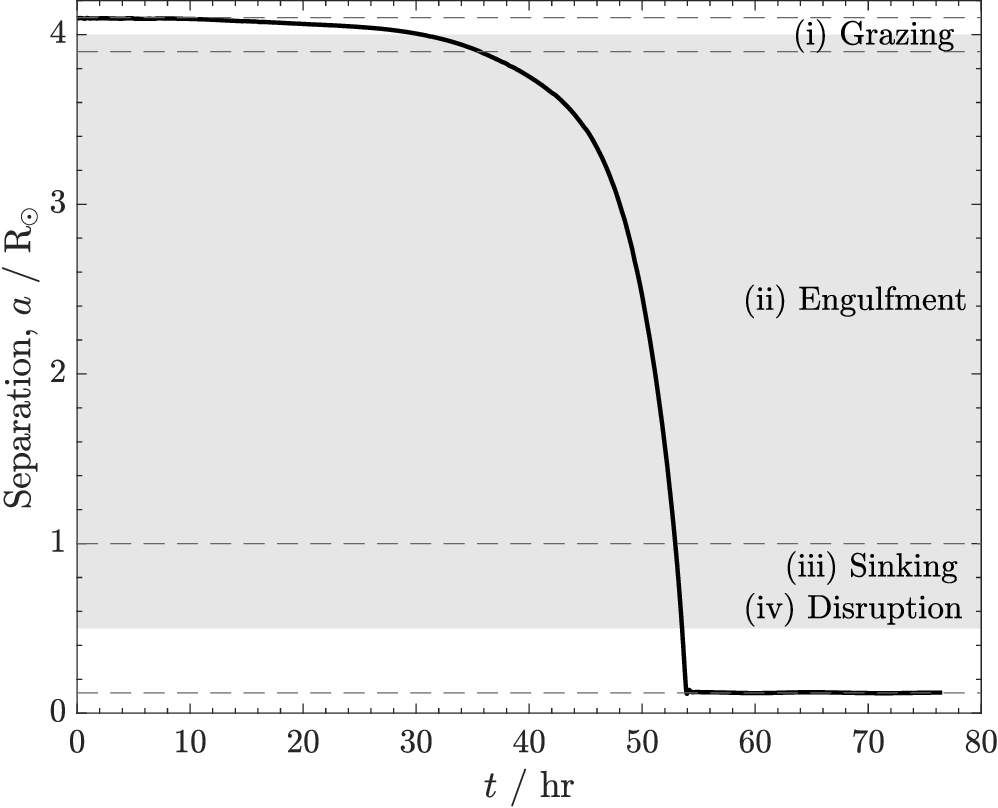}
    \caption{Evolution of the separation, $a$, between the planet and the stellar core. The horizontal dashed lines mark the boundaries of the grazing and engulfment phases of planetary engulfment that we have identified in Section \ref{subsec:phases}. The grey shaded region shows the extent of the convective envelope.}
    \label{fig:sep}
\end{figure}

We identify four different stages of planetary engulfment, similar to the processes described by \cite{Metzger+12} and \cite{Jia+Spruit18}. Figure \ref{fig:sep} plots the separation, $a$, between the stellar core and the planet as a function of time, with these four stages labelled. We define $a$ as the distance between the point-mass stellar core and the densest planet particle.

\paragraph*{(i) Grazing phase}
The \ac{HJ} experiences a relatively long `grazing phase' near the stellar surface (panels a)-b)). We label this phase in Figure \ref{fig:sep} as the part of the inspiral where the \ac{HJ} is partially immersed within the stellar envelope ($\Rstar-\Rp<a<\Rstar+\Rp$). This period lasts 35\hr, during which the \ac{HJ} completes more than one and a half orbits. Before this, the \ac{HJ} experiences an even longer period of orbital decay driven by tidal forces and atmospheric drag, which is beyond the scope of our simulation.

As explained in Section \ref{subsec:regimes}, hydrodynamical drag drives orbital decay in the simulated regime. In the stellar model, the density scale height near the surface, $H\sim 10^{-2}\Rsun$, is a small fraction of the planet radius. Yet, $H$ increases with depth, surpassing $\Rp = 0.1 \Rsun$ upon reaching $a=3.9\Rsun$, where the \ac{HJ} becomes fully immersed. Therefore, orbital decay during the grazing phase is driven by interaction with upstream gas that is within a distance $H$ from the planet's inner edge closest to the stellar centre \citep{Metzger+12}. The cross-sectional area of this portion of the planet is $\sim \Rp^{1/2}H^{3/2}$, and the density of the upstream gas that interacts with this region is approximately $\rho(r=a-\Rp)$. Therefore, the drag luminosity scales as $\dot{E}_\mathrm{hydro} \sim \Rp^{1/2}H^{3/2}\rho(r=a-\Rp)v^3$.

However, the resolution at the stellar surface limits our ability to accurately resolve the grazing phase. Initially, the surface of the 3D star is resolved down to $10^{-4}\gcc$, which is three orders of magnitude larger than the photospheric density according to the \MESA model (see also Appendix \ref{app:resolution}). The minimum resolved density scale height is $\approx 0.3\Rsun$, which is already three times the planet radius. The decay timescale, set by the time required to migrate inwards by a scale height from the stellar surface, scales as $H^{-1/2}\rho^{-1}$, implying that the drag-dominated part of the grazing phase in our simulations may be $10^3-10^4$ times shorter than in actuality. Local simulations \citep[e.g.][]{Sandquist+98,Sandquist+02,Yarza+22} or 1D studies \citep[e.g.][]{Soker+84,Livio&Soker84,Metzger+17,OConnor+23}, which may easily resolve the threshold density scale height, are better suited for studying the grazing interaction.

\paragraph*{(ii) Engulfment}
The main engulfment phase (panels b)-d)) follows the grazing phase. The \ac{HJ} spirals in from $3.9\Rsun$ to $1.0\Rsun$ in 18.5 hr (see Figure \ref{fig:sep}), approximately completing one orbit. The orbital decay timescale is therefore comparable with the orbital period ($a/(-\dot{a}) \sim P_\mathrm{orb}$). During this phase, the \ac{HJ} is completely immersed and the local scale height exceeds the planet radius, so that the interaction cross-section is the geometric cross-section, $\pi\Rp^2$. $\Rp/H$ ranges from a few times $10^{-1}$ early on to a few times $10^{-2}$ deeper within the convective envelope. This means that in the early inspiral, the density gradient of the upstream flow may have a dynamically important effect. This may be modelled as a modification to the drag coefficient, which has been studied extensively in `wind tunnel' simulations \citep[e.g.][]{MacLeod&Ramirez-Ruiz15,MacLeod+2017,De+20,Yarza+22}. The \ac{HJ}'s instantaneous velocity is close to but slightly less than the local Keplerian speed, $\vKep = (Gm/a)^{1/2}$ (see Section \ref{subsec:vel}), where $m$ is the stellar mass coordinate at the planet's position. The orbital motion is mildly supersonic relative to the envelope, with $\mathcal{M} \approx 3$, driving a bow shock through the stellar envelope (Figure \ref{fig:rho_xy}c)-e)). This can be seen in the inset of Figure \ref{fig:rho_xy}d), which plots the velocity divergence, $\nabla\cdot\mathbf{v}$. The bow shock can be seen as the dark blue band surrounding the planet, which is followed by a red post-shock expansion region. In this phase, the \ac{HJ}'s mass steadily decreases by $\approx 90\%$, down to 0.1\Mj. The ablated material lies in the planet's wake, as highlighted by Figure \ref{fig:rho_planet}. The boundary between this material and the stellar gas is susceptible to the Kelvin-Helmholtz instability, which leads to the growth of vortices seen in panel f). We discuss the \ac{HJ}'s ablation process in Section \ref{subsec:ablation}.

\paragraph*{(iii) Sinking}
The sinking phase is characterised by the transition from azimuthal orbital motion to a radial plunge (panels d)-e)). In Figure \ref{fig:sep}, we mark the start of this phase as the point in time when the inward radial velocity exceeds the azimuthal velocity, $-v_r > v_\phi$. We reiterate the result from Section \ref{subsec:regimes} that radial descent may only occur in engulfment events driven by hydrodynamical drag, but not gravitational drag, since in the latter regime, the inspiral timescale is a factor $\Mstar/\Mp \gg 1$ longer than the orbital period at contact (following equation (\ref{eq:tBHL})). We discuss the transition to sinking in greater detail in Section \ref{subsec:vel}.

\paragraph*{(iv) Disruption}
The remaining \ac{HJ} is eventually disrupted near the inner boundary of the convective envelope due to the large stellar densities encountered there, which are conducive to both tidal and ram-pressure disruption. We stop the simulation after the disrupted material approaches rest, at $t=76.6\hr$ (panel f)), although there is still significant rotational motion near the stellar surface from ablated material and angular momentum deposited during the grazing phase (see Section \ref{subsec:spinup}). Figure \ref{fig:sep} shows that the deepest material settles at 0.1\Rsun, penetrating into the radiative region but not beneath the simulation's inner boundary (i.e. above the H-burning shell). We further discuss the disruption mechanism and the maximum depth reached by the disrupted material in Section \ref{subsec:disruption}.



\subsection{Planet velocity}
\label{subsec:vel}
\begin{figure}
    \includegraphics[width=\linewidth]{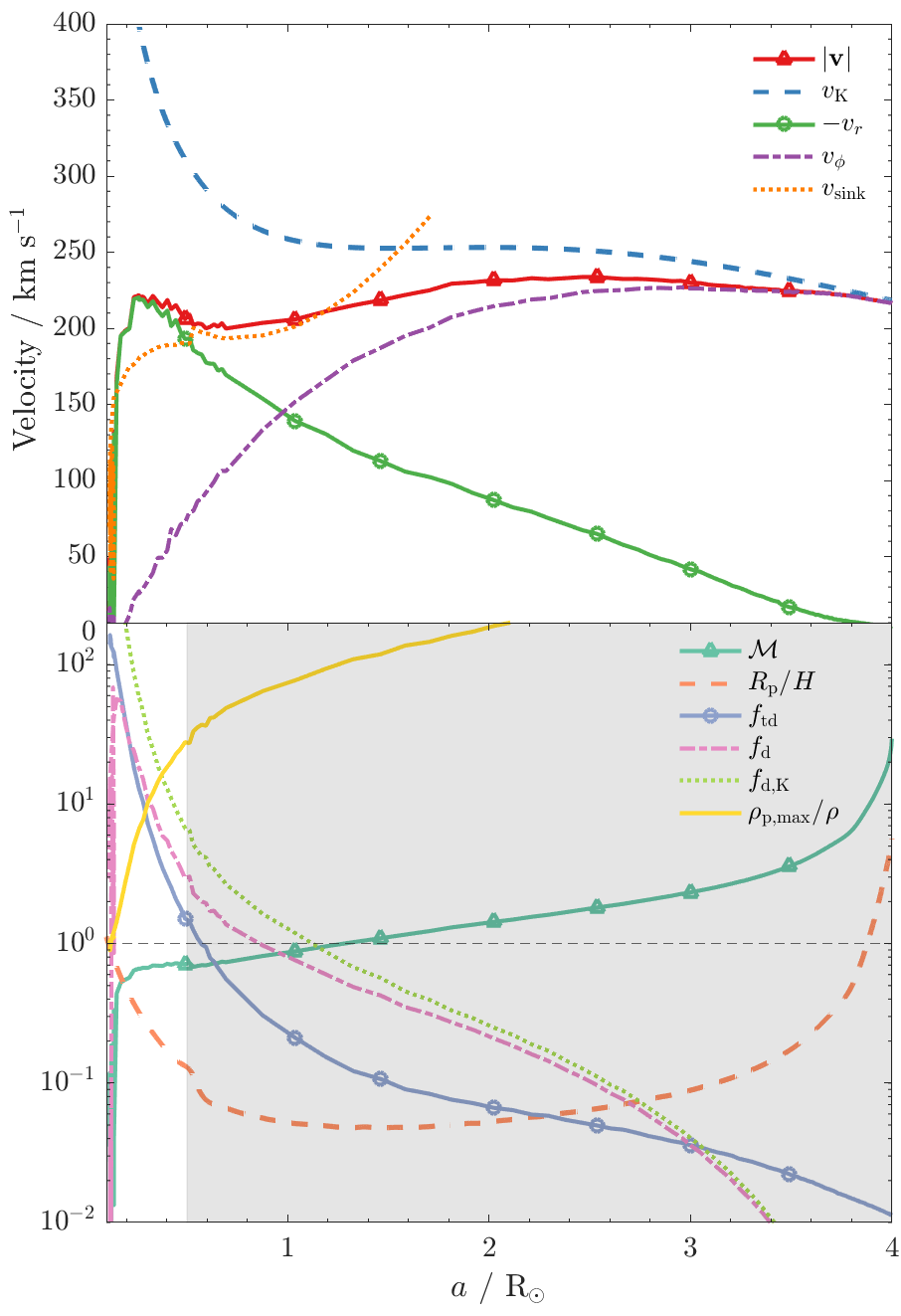}
    \caption{Evolution of key parameters as functions of the planet separation. Top panel: Planet velocity components, including the planet speed ($|\mathbf{v}|$), (negative) radial velocity ($-v_r$), and azimuthal velocity ($v_\phi$). For comparison, we also show the local Keplerian speed (\vKep) and the descent speed ($v_\mathrm{sink}$) given by equation (\ref{eq:vsink}). Bottom panel: Key dimensionless quantities that characterise various processes, including the planet's Mach number ($\mathcal{M}$), the ratio of planet radius to density scale height ($\Rp/H$), tidal disruption parameter ($f_\mathrm{td}$, equation (\ref{eq:td})), ram-pressure disruption parameter ($f_\mathrm{d}$, equation (\ref{eq:disruption})), $f_\mathrm{d}$ evaluated with the Keplerian speed ($f_\mathrm{d,K}$), and the ratio of the maximum density inside the planet to the background density ($\rho_\mathrm{p,max}/\rho$). The grey region shows the extent of the convective envelope.}
    \label{fig:vel}
\end{figure}

The top panel of Figure \ref{fig:vel} shows components of the \ac{HJ}'s velocity, $\mathbf{v}$, as functions of separation. As with the position, we define $\mathbf{v}$ as the velocity of the densest \SPH particle that initially makes up the \ac{HJ}. The planet's speed, $|\mathbf{v}|$ (red curve with triangle markers), is between $\approx 200-240\kms$ up to the last $\approx 0.2\Rsun$, where the planet is brought to rest. Because $|\mathbf{v}|$ remains mostly constant despite interacting with deeper material with higher sound speeds, the Mach number gradually decreases from $\approx 10$ at the start of the engulfment phase to 0.8 upon reaching the sinking phase (bottom panel of Figure \ref{fig:vel}). The local Keplerian speed, $\vKep$ (blue dashed curve), is a reasonable approximation to $|\mathbf{v}|$ until radial motion dominates, after which it is no longer relevant.

The transition to sinking is shown by the increasing radial ($-v_r$, green curve with circle markers) and diminishing azimuthal ($v_\phi$, purple dot-dashed curve) velocity components. At $a=1\Rsun$, $-v_r$ exceeds $v_\phi$, and for $a\lesssim0.2\Rsun$, $|\mathbf{v}|$ is entirely radial and rapidly decreases in magnitude. We note that the noisiness in these plots at small separations reflects sensitivity to the choice of \SPH particle used to define $\mathbf{v}$.

Neglecting the planet's inertia during the sinking phase, the radial descent speed is determined by the balance between hydrodynamical drag and gravity \citep{Jia+Spruit18}:
\begin{align}
    v_\mathrm{sink} = \biggl( \frac{2Gm\Mp}{\pi\Rp^2\rho a^2} \biggr)^{1/2} = \biggl(\frac{2\Mp}{\pi\rho \Rp^3}\frac{\Rp}{a}\biggr)^{1/2}\vKep(a),
    \label{eq:vsink}
\end{align}
where we have preserved the factor of $\pi$ in the \ac{HJ}'s geometric cross-section and assumed a drag coefficient of 1/2 appropriate for a smooth sphere at large Reynolds numbers ($Re\gtrsim 10^4$). We have also neglected buoyancy, which is appropriate at least within the convective region, where the mean planet density, $\langle\rho_\mathrm{p}\rangle \approx 1.4\gcc$, is many times greater than the ambient density. We plot $v_\mathrm{sink}$ as the orange dotted curve in the top panel of Figure \ref{fig:vel} for $a<1.7\Rsun$. We used the \ac{HJ}'s instantaneous mass for $\Mp$ (see Section \ref{subsec:ablation}), a fixed radius $\Rp = 0.1\Rsun$, and assumed the unperturbed stellar density, $\rho(r=a)$, at the present separation. We find that $v_\mathrm{sink}$ agrees reasonably with $|\mathbf{v}|$ from $a\lesssim 1.2\Rsun$, despite the sinking phase only starting at 1.0\Rsun. The sharp bump at $a=0.5\Rsun$ arises from the density bump at the base of the convective zone. At small separations ($a\lesssim 0.5\Rsun$), $v_\mathrm{sink}$ under-predicts the actual speed. This could be because the \ac{HJ} radius significantly decreases below the assumed 0.1\Rsun, and because of rapid mass loss in these dense stellar layers, which could impart a large acceleration onto the planet not accounted for in equation (\ref{eq:vsink}).

\subsection{Planet ablation} \label{subsec:ablation}

\begin{figure}
    \includegraphics[width=\linewidth]{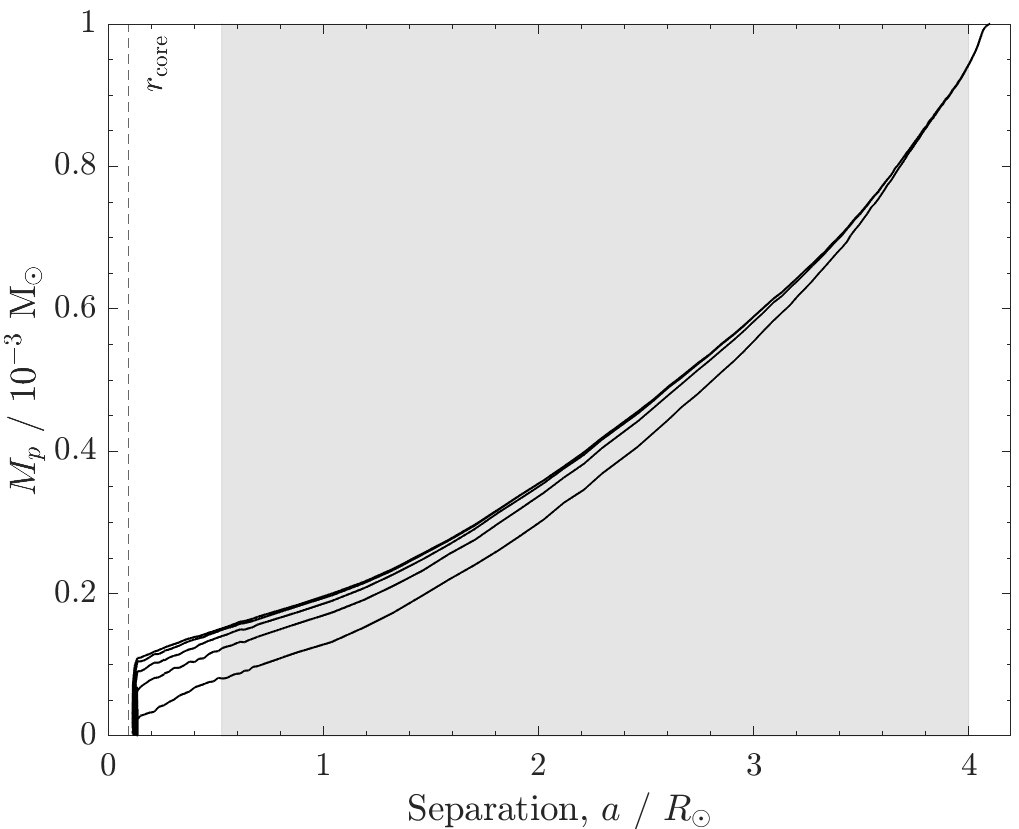}
    \caption{Evolution of the planet mass, \Mp, as a function of the separation, $a$, showing that the hot Jupiter loses around 90\% of its mass before disrupting. Around $0.9\Mj$ of material is deposited into the convective zone (grey region). The five different lines estimate \Mp according to the different velocity thresholds explained in Section \ref{subsec:ablation}.}
    \label{fig:mass}
\end{figure}

\begin{figure*}
    \includegraphics[width=\linewidth]{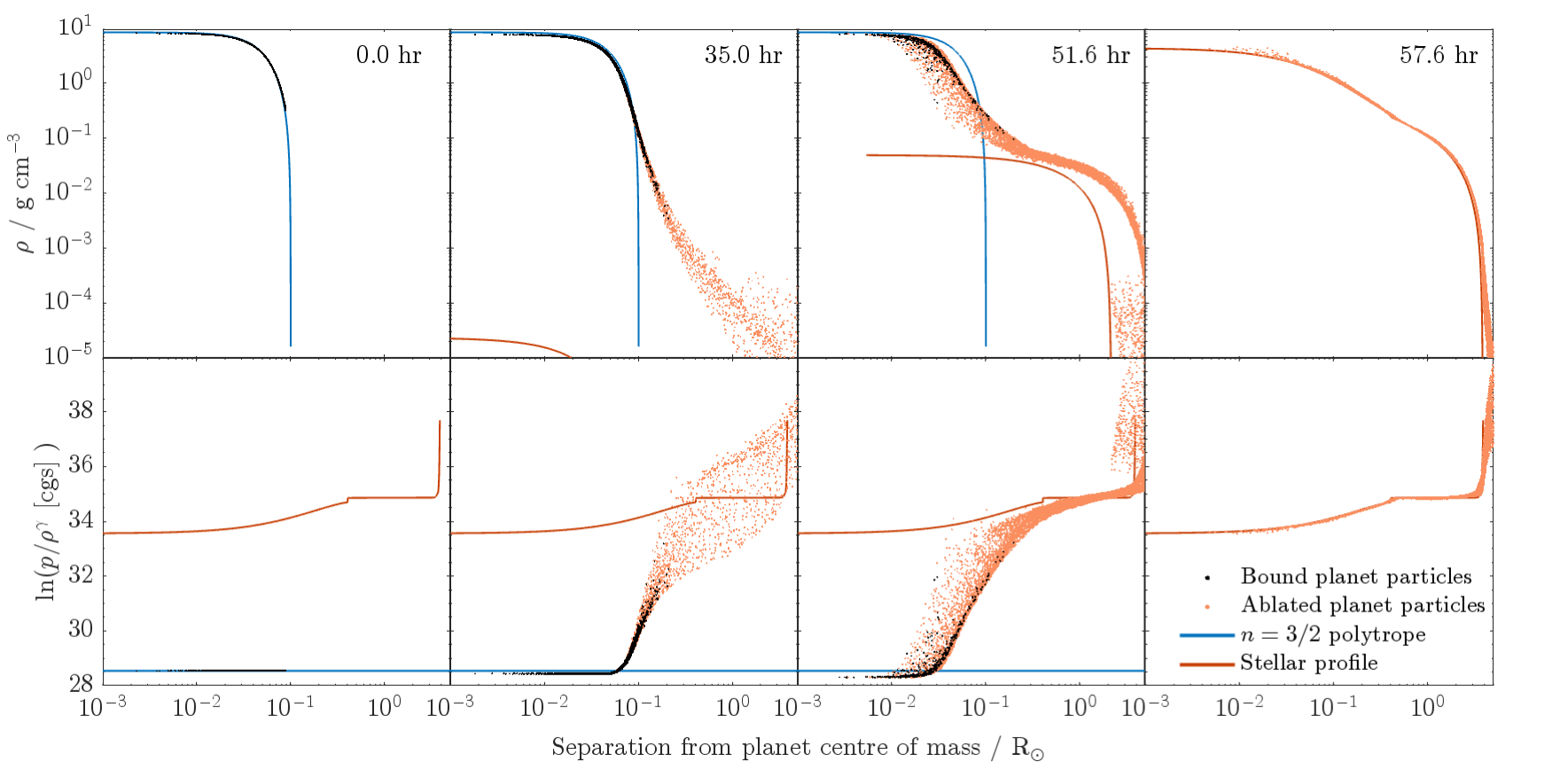}
    \caption{\ac{HJ}'s radial structure at four different points in the simulation, with the first three columns corresponding to panels a), b), and d) in Figure \ref{fig:rho_xy}, while the rightmost column occurs shortly after planet disruption. Top panels: Density profile. Bottom panels: Entropy, defined as $\ln(p/\rho^\gamma)$ where $p$ and $\rho$ are pressure and density in cgs units, and $\gamma=5/3$ is the adiabatic exponent. Bound planet \SPH particles are plotted as black dots, while ablated planetary material is shown as orange dots. The $n=3/2$ polytrope used for modelling the \ac{HJ} is in blue. The red line shows a shifted stellar profile, plotted from the planet's current centre of mass to show the local conditions encountered by the planetary material. In the rightmost column, the origin is taken to be the stellar centre since the \ac{HJ} has completely disrupted.}
    \label{fig:planet_rho_s_r}
\end{figure*}

Figure \ref{fig:mass} shows the evolution in planet mass, \Mp, as a function of the separation, $a$. We compute \Mp by summing the masses of \SPH particles that (i) initially made up the \ac{HJ}, (ii) have velocity projections that are within some threshold of the velocity at the densest point of the \ac{HJ}, and (iii) are within 2\Rj from that densest point. The five different lines in Figure \ref{fig:mass}, from bottom to top, assume velocity thresholds of 0.1, 0.3, 0.5, 0.7, and 0.9 (most to least restrictive). We assume a threshold of 0.7 in the rest of the discussion.

The planet loses around 90\% of its mass during the engulfment process. The mass-loss rate ranges from $\approx 10^{-5}\Msun~\mathrm{hr}^{-1}$ in the outer regions of the envelope to $\approx 10^{-4}\Msun~\mathrm{hr}^{-1}$ near the middle of the convective zone. However, the exact ablation rate at a given time or separation has not fully converged (see Appendix \ref{app:resolution}). The initial mass loss taking place in low-density surface layers is overestimated at lower resolutions. Despite this, the total ablated mass appears to be relatively robust, that is, later mass loss occurring in deeper layers adjusts to strip the \ac{HJ} by a similar overall amount. Across all tested resolutions, roughly 0.9\Mj of material is deposited into the convective zone, shown as the shaded region in Figure \ref{fig:mass}.

In an upcoming work, we present simulations that resolve planet ablation on a much finer scale using a wind tunnel setup, confirming that the observed ablation rates in the convective envelope agree with converged results within a factor of a few (Lau et al. in prep.), and that ablation in denser layers is closer to convergence. The result that the \ac{HJ} loses most of its mass during the engulfment process is consistent with \cite{Sandquist+98}, who simulated \ac{HJ} engulfment by a main-sequence star. \cite{Abia+20} and \cite{Cabezon+23} conducted global \SPH simulations of brown dwarf engulfment, also finding ablation but concentrated at very large depths. This could be because their brown dwarf model is around ten times more massive than our \ac{HJ}, and therefore more compact given similar radii.

Figure \ref{fig:planet_rho_s_r} shows the \ac{HJ}'s density and entropy profile at four different points in the simulation. The first three columns correspond to panels a), b), and d) in Figure \ref{fig:rho_xy}, while the fourth column shows the profile shortly after planet disruption. The black dots, representing \SPH particles that currently make up the planet, initially follow the $n=3/2$ polytrope used for modelling the \ac{HJ}, which has uniform entropy. The second and third columns show the planet's readjustment to mass loss and compression by the stellar gas on a dynamical timescale. The material ablated into the stellar envelope follows the density and entropy of the stellar profile (red line). The last column shows that after complete disruption (see Section \ref{subsec:disruption}), the ablated planetary material (orange dots) settles at a depth determined by density and entropy equilibrium with its surroundings.

The nearly factor of ten change in planet mass observed in our simulations cautions against assuming \Mp to be fixed during planetary engulfment, whether for analytical estimates \citep[e.g.][]{MacLeod+18,Stephan+18,Stephan+20,Soares-Furtado+21} or as a simplifying assumption in calculations \citep[e.g.][]{Staff+16,Yarza+22}. That mass loss occurs gradually and not as a single, global episode \citep{Jia+Spruit18} could impact the ability to enhance the surface chemical abundance of stars (see Sections \ref{subsec:disruption} and \ref{subsec:enrichment}).

A number of works have indicated that the mass ablation rate may scale
with the rate at which linear momentum is intercepted by the planet \citep[e.g.][]{Cheng74,Jia+Spruit18,Hirai+2020}. Our upcoming simulations also confirm this (Lau et al. in prep.). The \ac{HJ} intercepts momentum at the rate $\dot{p} \sim \pi \Rp^2 \rho |\mathbf{v}|^2$. On the other hand, ablation removes momentum at a minimum rate of $\dot{M}_\mathrm{p} v_\mathrm{esc}$, where $v_\mathrm{esc,p}=(2G\Mp/\Rp)^{1/2}$ is the escape speed at the planet's surface. Equating the two rates and using values for the density and velocity recorded at $a=2.8\Rsun$, we obtain
\begin{align}
    -\dot{M}_\mathrm{p} \lesssim \frac{\dot{p}}{v_\mathrm{esc}} \sim 3\times10^{-4} \Msun~\mathrm{hr}^{-1}
    \bigg( \frac{\rho}{0.012 \gcc} \bigg)
    \bigg( \frac{|\mathbf{v}|}{230\kms} \bigg)^2.
    \label{eq:Mablated}
\end{align}
This estimate is only an upper limit because momentum is also expended in producing ram-pressure drag and because material could be leaving the planet faster than $v_\mathrm{esc}$. It also neglects other processes that influence mass loss, including the \ac{HJ}'s adiabatic expansion in response to mass loss and compression by ram pressure and ambient pressure. Nonetheless, this crude estimate shows that the mass-loss rates observed in the simulations are plausible.

Mass loss may also occur via thermal evaporation, where the thermal energy of the background medium overcomes the \ac{HJ}'s binding energy. This process is not modelled in our simulation, which has no heat transport except for resolution-scale conductivity, and therefore requires separate discussion. Thermal evaporation occurs when the ambient temperature or temperature of the shocked gas around a supersonic planet exceeds the \ac{HJ}'s virial temperature \citep{Livio&Soker84,Siess&Livio99a,Aguilera-Gomez+16},
\begin{align}
    T_\mathrm{virial} \sim \frac{G\Mp m_\mathrm{H}}{k_\mathrm{B}\Rp}
    \approx 2\times10^5~\mathrm{K}~\biggl( \frac{\Mp}{\Mj} \biggr) \biggl(\frac{\Rp}{\Rj}\biggr)^{-1},
\end{align}
where $m_\mathrm{H}$ is the proton mass and $k_\mathrm{B}$ is Boltzmann's constant. For Jovian planets, $T_\mathrm{virial}$ greatly exceeds the surface temperature of any host star, meaning thermal evaporation is unimportant during and leading to the grazing phase\footnote{Although, the shortest-period \acp{HJ} are known to have inflated radii of up to 2\Rj, correlated with the level of irradiation \citep{Lopez&Fortney16}.}. Inside a giant star's convective envelope, the temperature easily exceeds $T_\mathrm{virial}$, especially after shock passage. A key question is whether the duration of the \ac{HJ}'s dynamical in-fall exceeds the heating timescale. While giant planets are expected to be largely if not fully convective \citep{Guillot+04}, convection cannot transport energy inwards against both a temperature inversion and a chemical gradient set by the higher mean molecular weight of the planetary material. Electron thermal conduction is only efficient near the degenerate core, and is unlikely to be important in the bulk of the envelope \citep{Hubbard68,Stevenson+Salpeter77}. This leaves radiative diffusion as the remaining heating mechanism. As a rough estimate, the photon diffusion timescale is
\begin{align}
  t_\mathrm{diff} \sim \frac{\kappa \rho \Rp^2}{c} = 20~\mathrm{d}\bigg(\frac{\kappa}{10^{-2}~\mathrm{cm}^2~\mathrm{g}^{-1}}\bigg) \bigg(\frac{\rho}{0.1\gcc}\bigg),
\end{align}
assuming a gas line opacity of $\sim 10^{-2}~\mathrm{cm}^2~\mathrm{g}^{-1}$ \citep{Stevenson91}. This exceeds the duration of the inspiral, and the actual thermal timescale is much longer by a factor equal to the ratio of gas to radiation energy. Thermal evaporation is therefore unlikely to be important.


\subsection{Disruption and penetration depth} \label{subsec:disruption}
The \ac{HJ} is expected to disrupt at sufficiently large depths. We discuss the expected location and possible mechanisms of disruption. \cite{Jia+Spruit18} estimate that an engulfed planet dissociates in a global process when the ram pressure overcomes its binding energy. This roughly occurs when the dimensionless disruption parameter, $f_\mathrm{d}$, exceeds unity:
\begin{align}
    f_\mathrm{d} \equiv \frac{\rho \Delta v^2}{\langle\rho_\mathrm{p}\rangle v_\mathrm{esc,p}^2} \gtrsim 1,
    \label{eq:disruption}
\end{align}
where $\rho$ is the ambient density, $\Delta v$ is the velocity contrast between the planet and the upstream gas, and $\langle\rho_\mathrm{p}\rangle = 3\Mp/(4\pi\Rp^3)$ is the mean planet density. We have shown that during the grazing and engulfment phases, $\Delta v$ may be reasonably approximated by the local Keplerian speed, $\vKep$, in which case the disruption parameter may be written as
\begin{align}
    f_\mathrm{d,K} = \frac{\rho}{2\langle\rho_\mathrm{p}\rangle} \frac{m}{\Mp} \frac{\Rp}{a}.
\end{align}
Tidal forces in the stellar interior may also cause global disruption, which occurs when the volume-averaged stellar density interior to the planet exceeds the mean planet density. That is, the following tidal disruption parameter exceeds unity:
\begin{align}
    f_\mathrm{td} \equiv \frac{m}{\Mp} \bigg(\frac{\Rp}{a}\bigg)^3 \gtrsim 1.
    \label{eq:td}
\end{align}

The bottom panel of Figure \ref{fig:vel} shows $f_\mathrm{td}$, $f_\mathrm{d}$, and $f_\mathrm{d,K}$ as functions of $a$. In calculating their values in the simulation, we assume a fixed radius of $\Rp = 0.1\Rsun$ but use the instantaneous mass \Mp calculated in Figure \ref{fig:mass}. We also assume $\rho = \rho(r=a)$ is given by the unperturbed stellar density at the \ac{HJ}'s position. The parameter $f_\mathrm{d}$, shown by the pink curve, exceeds unity at $a = 0.83\Rsun$, within the convective region and near the start of the sinking phase. Because the Keplerian speed overestimates the actual planet speed, and therefore the ram pressure of the incident flow, $f_\mathrm{d,K}$ reaches unity at a larger radius, $a = 1.1\Rsun$. On the other hand, $f_\mathrm{td}$ exceeds unity at $a=0.57\Rsun$, which is near but above the convective boundary. The criteria for ram pressure disruption are therefore met before the criterion for tidal disruption. However, because equations (\ref{eq:disruption})-(\ref{eq:td}) only provide approximate scalings, the locations at which ram-pressure or tidal destruction occur are only known roughly. Also, the scale heights of $f_\mathrm{d}$ and $f_\mathrm{td}$ at the boundaries where they exceed unity, the distance between the two boundaries, and the planet size itself are all comparable. As such, both tides and ram pressure are likely to have played a role in disrupting the \ac{HJ} near $a\approx 0.6-0.8\Rsun$.

The disrupted planetary material, which is denser than its surroundings, penetrates into deeper layers of the star. Figure \ref{fig:sep} shows that the densest material reaches $\approx0.1\Rsun$, which is within the radiative shell. These fragments stall when the density contrast reduces to zero, upon which they become neutrally buoyant \citep[equivalently, they come into entropy equilibrium with their surroundings, e.g.][]{Ivanova+02}. This is illustrated in the rightmost panels of Figure \ref{fig:planet_rho_s_r}, and also by plotting the ratio $\rho_\mathrm{p,max}/\rho$ of the maximum planet density to the ambient density, shown as the yellow curve in the bottom panel of Figure \ref{fig:vel}. Because $\rho_\mathrm{p,max}$ does not vary significantly over the course of the spiral-in, the decrease in $\rho_\mathrm{p,max}/\rho$ is driven almost entirely by the rise in ambient density. The descent of the densest planetary material halts where the ratio reaches one, at $a=0.12\Rsun$. The material that has been ablated in the convective envelope will be mixed over many convective turnover times, which is $\sim100$ days near the base of the convective zone. The deepest material, having penetrated into the radiative layer but having a larger mean molecular weight than its surroundings, is unstable against thermohaline convection, which may mix the material even deeper on the thermal timescale \citep{Vauclair04}. Our simulation is incapable of capturing these processes as we do not model radiative and chemical diffusion.

\section{Discussion} \label{sec:discussion}
\subsection{Chemical enrichment signature}
\label{subsec:enrichment}
The amount of ablated planet mass in our simulations can be used to predict the surface chemical enrichment signature produced in the host star. Numerous studies have focused on the enrichment of \Li, which is a sensitive measure of stellar evolution. Canonical stellar evolutionary models predict that \Li is gradually depleted during red giant evolution, as the deepening of the convective region during the first dredge-up mixes \Li-depleted material to the surface. Yet, $\approx 1.2\%$ of giant stars are \Li-rich ($\ALi > 1.5$) \citep[e.g.][]{Wallerstein&Sneden82,Balachandran+00,Drake+02,Ruchti+11,Monaco+11,Martell&Shetrone13,Casey+16,Kirby+16,Casey+19,Gao+19}, prompting suggestions that planets, whose interiors are too cool to burn \Li, could deposit \Li-rich material into their hosts \citep{Aguilera-Gomez+16,Soares-Furtado+21}.

Mixing $N_\mathrm{p}$ \Li atoms into a convective envelope initially containing $N_\star$ \Li atoms increases \ALi by $\Delta \ALi = \log_{10}( 1+N_\mathrm{p}/N_\star )$. We assume our red giant model has baseline \Li abundance of $\ALi = 0.85$, which is consistent with data compiled by \cite{Soares-Furtado+21} from the GALAH (GALactic Archaeology with HERMES) survey. We assume that the \ac{HJ} has meteoritic \Li abundance of $\ALi = 3.30$ \citep{Montalban&Rebolo02}, and therefore $10^{3.30-0.85} \approx 280$ times more \Li atoms per hydrogen atom than the stellar gas. This is diluted by the ratio of the ablated planet mass ($9\times10^{-4}\Msun$) to the convective envelope mass (0.75\Msun), resulting in $\Delta \ALi = \log_{10}[1+280(9\times10^{-4} / 0.75)] \approx 0.13$.

This enhancement signature is not expected to be statistically significant, as it is comparable with the typical \ALi spread in cluster red giants of $\approx 0.1$ \citep{Lind+09,Romano+21,Soares-Furtado+21}. \cite{Soares-Furtado+21} estimate that the signature may persist for a few times $10^7\yr$ on the lower red giant branch, although they do not model uncertain processes such as thermohaline mixing, elemental diffusion, and convective overshoot \citep[see][]{Theado&Vauclair12,Sevilla+22}. \ac{HJ} engulfment by a more evolved host star would produce a stronger enrichment signature due to its lower baseline \Li abundance and less massive convective zone. However, the planet would also interact with slower and less dense gas, which could reduce the amount of surface pollution unless complete tidal disruption occurs in the convective envelope. Future studies can provide a more complete understanding by calculating the \ac{HJ} mass ablation across host stars at different points in their evolution.

\subsection{Induced stellar rotation}
\label{subsec:spinup}
\begin{figure}
    \includegraphics[width=\linewidth]{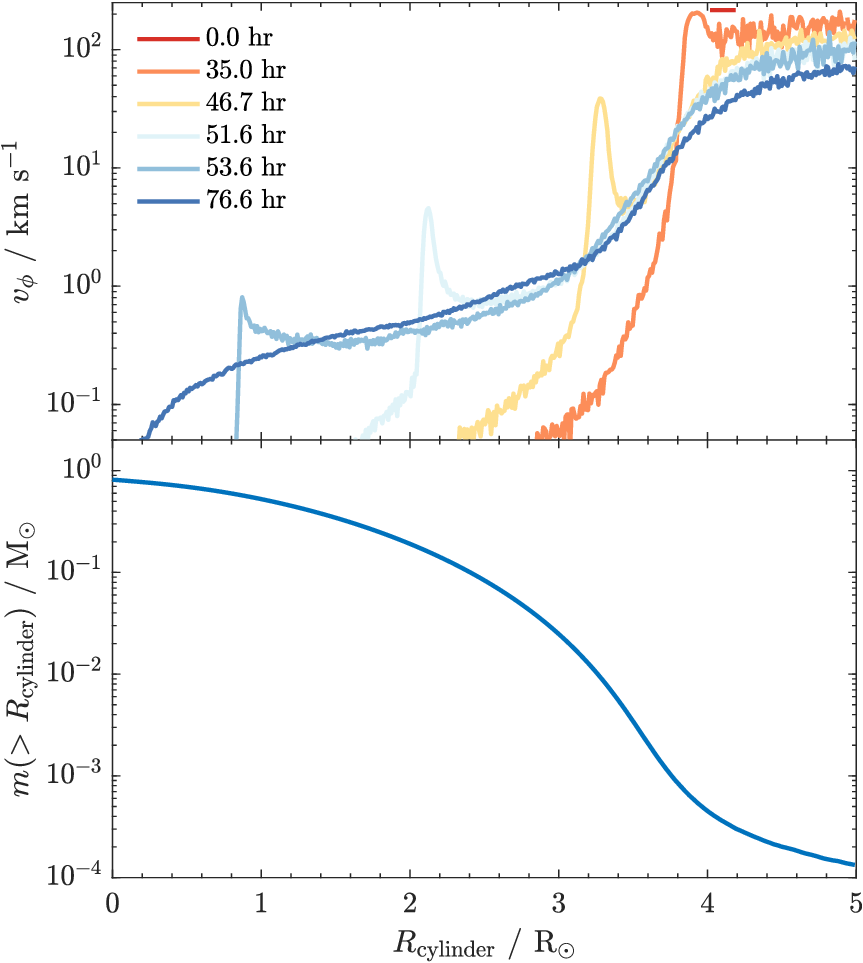}
    \caption{Top panel: Azimuthal velocity, $v_\phi$, as a function of cylindrical radius, $R_\mathrm{cylinder}$, at different points in time corresponding to the snapshots in Figure \ref{fig:rho_xy}. Bottom panel: Mass exterior to $R_\mathrm{cylinder}$. At $t = 0$, there is no rotation other than the orbital motion of the hot Jupiter itself, which is shown as a flat red line at the Keplerian velocity, 216\kms. The inward movement of the hot Jupiter is shown by the spike. In the last simulation snapshot, the outermost $\approx 10^{-3}\Msun$ of material rotates at a few times $10\kms$.}
    \label{fig:vphi_profile}
\end{figure}

Planetary engulfment may impart significant rotation to the host star by transferring the planet's orbital angular momentum to the stellar envelope. This has been suggested to account for observations of some rapidly rotating giant stars \citep[e.g.][]{Soker&Harpaz00,Carlberg+13,Privitera+16b}. We examine how planetary engulfment has impacted the azimuthal velocity profile, $v_\phi$, as a function of cylindrical radius, $R_\mathrm{cylinder}$, in Figure \ref{fig:vphi_profile}. The top panel plots azimuthally averaged $v_\phi$ profiles for the snapshots shown in Figures \ref{fig:rho_xy} and \ref{fig:rho_planet}. At $t=0$, there is no rotation other than the orbital motion of the \ac{HJ} itself, shown as the flat red segment at $4.0 < R_\mathrm{cylinder} / \Rsun< 4.2$ with the local Keplerian velocity, 216\kms. The grazing interaction causes a low-density surface layer to expand beyond the initial stellar radius of 4\Rsun. This material spins up to 100\kms, although representing a very small amount of mass. This is illustrated by the bottom panel, which plots the mass exterior to a given value of $R_\mathrm{cylinder}$. By the end of the simulation (dark blue curve), only the outermost $\sim10^{-3}\Msun$ of material rotates faster than 10\kms. The bulk of the stellar envelope is slowly rotating ($v_\phi \sim 1\kms$), in agreement with the simulations by \cite{Staff+16} for red giant and asymptotic giant branch stars. Furthermore, this angular momentum will be redistributed on a much longer timescale than simulated. Modelling this process is beyond the scope and capabilities of simple adiabatic simulations, which neglect, for example, convection and magnetic fields. But it is straightforward to estimate the maximum spin-up by assuming angular momentum conservation as follows.

The induced velocities are slightly faster than the envelope's convective eddies. That is, the envelope has a small convective Rossby number ($Ro\lesssim 1$), which in this case is defined as the ratio of the \BV frequency to the imparted angular frequency. In the small $Ro$ regime, angular momentum is redistributed to produce rigid rotation. Assuming the orbital angular momentum is perfectly transferred into the convective envelope's rotation,
\begin{align}
    I_\mathrm{env}\Omega_\mathrm{env} = \Mp\Rstar^2\Omega_\mathrm{orb}.
\end{align}
Here, $I_\mathrm{env}$ is the convective region's moment of inertia, assumed to be unaltered by the engulfment process for a small mass ratio. Engulfing more massive planets or brown dwarfs is likely to inflate a more evolved host star, significantly increasing its moment of inertia before contracting on the thermal timescale. $\Omega_\mathrm{env}$ is the angular velocity the envelope rotates with as a rigid body after engulfment. The right-hand side is the planet's orbital angular momentum upon entering the host star on a circular orbit, with angular velocity $\Omega_\mathrm{orb} = (G\Mstar/\Rstar^3)^{1/2}$. The increase in surface rotational velocity, $\Delta v_\mathrm{rot}=\Omega_\mathrm{env} R_\star$, can then be solved, giving
\begin{align}
    \Delta v_\mathrm{rot}
    \approx 1.7 \kms \biggl( \frac{k_\mathrm{env}}{0.13} \biggr)^{-1}
    \biggl( \frac{M_p/M_\star}{10^{-3}} \biggr)
    \biggl( \frac{M_\star}{1\Msun} \biggr)^{1/2}
    \biggl( \frac{R_\star}{4\Rsun} \biggr)^{-1/2},
    \label{eq:vrot}
\end{align}
where we have expressed the convective envelope's moment of inertia in terms of an apsidal motion constant $k_\mathrm{env}$, defined via $I_\mathrm{env} = k_\mathrm{env} \Mstar \Rstar^2$. $k_\mathrm{env}$ is generally a quantity of the order of 0.1 for evolved giant stars, and decreases with age as the star becomes more centrally concentrated. For our stellar model, $k_\mathrm{env} = 0.13$, which yields $\Delta v_\mathrm{rot} = 1.7\kms$ according to equation (\ref{eq:vrot}).

Our estimate is roughly consistent with the velocity magnitudes observed in the bulk of the envelope ($R_\mathrm{cylinder} \lesssim 3.5\Rsun$) in the last profile shown in Figure \ref{fig:vphi_profile}. The linear scaling of $\Delta v_\mathrm{rot}$ with \Mp implies that engulfing a 10\Mj companion could produce envelope rotation velocities up to 17\kms, which are seen in the simulations by \cite{Cabezon+23}. Recently, the red giant KIC 9267654 was reported to have a surface that rotates more rapidly than the bulk of its envelope \citep{Tayar+22}, which led to the suggestion that it had recently ingested a planet. Our simulation suggests that planetary engulfment could account for this non-monotonic rotation profile, although the duration of this state depends on the uncertain efficiency of angular momentum transport. This could be studied by mapping the rotation profile obtained at the end of our simulation into a detailed stellar evolution code such as \MESA.

\subsection{Luminous transients}
\label{subsec:transients}

Engulfment occurring on the subgiant and on the lower red giant branch can produce a luminous transient powered by orbital decay \citep{MacLeod+18}. During the early grazing phase, energy is deposited in surface layers and quickly radiated away before being able to thermalise, causing the observed luminosity to directly track the drag luminosity. Our simulation does not capture this phase as it requires resolving the low-density photosphere and modelling radiation transport. It can instead be studied using one-dimensional calculations that self-consistently model drag heating and energy transport in these layers \citep{OConnor+23}. 

During later parts of grazing and the transition to full engulfment, the injected orbital energy may be thermalised in the envelope and increases the stellar luminosity once it is transported to the photosphere. So a slow luminosity increase is expected during the grazing phase, followed by a sharp decline once the energy transport timescale exceeds the orbital decay timescale. \cite{Yarza+22} predict that \acp{HJ} engulfed on the red giant branch at 10\Rsun separations could brighten the star by factors of several hundred for a year, or up to 5,000\yr if engulfing a 50\Mj brown dwarf. The stellar luminosity then returns to pre-engulfment values on the envelope's thermal timescale. This secular dimming might have been observed in KIC 8462852 \citep{Metzger+17}.

Another source of photometric variability can be provided by the expansion, cooling, and recombination of material ejected from the engulfment process. In our model, about $1.6\times 10^{-5}\Msun$ of material becomes unbound. While the exact value has not converged with resolution, its order of magnitude is fairly robust (see Appendix \ref{app:resolution}). Applying the scaling relation of \citet{Matsumoto:2022} and assuming that $10^{-5}\Msun$ of material escapes at 300\kms (roughly the surface escape velocity) gives a peak luminosity of about $2\times10^{35}~\mathrm{erg~s}^{-1}$, almost ten times the unperturbed stellar luminosity, and a plateau duration of about 3 days. The estimate may break down if the ejecta are radiation pressure dominated or deviate from spherical geometry.

It is then possible that some planetary engulfments might exhibit lightcurves with two peaks: the first associated with drag-heating of the near-surface stellar layers, and the second due to H-recombination in ejected material. Their relative importance depends on the host star's evolutionary phase. For planetary engulfment occurring further up the red giant branch, the surface drag luminosity cannot surpass the background stellar luminosity \citep{MacLeod+18}, and so H-recombination might be the only source of photometric variability. At the same time, these evolved red giants have more loosely bound envelopes, and so may produce more unbound ejecta in an engulfment episode. The `sub-luminous red nova' ZTF SLRN-2020 was reported as a candidate for a recombination-powered transient due to planetary engulfment \citep{De+23}. This event has an optical luminosity of $\sim10^{35}~\mathrm{erg~s}^{-1}$ and was accompanied by mid-infrared emission. Its inferred progenitor parameters ($M_\star \sim 0.8-1.5\Msun$, $R_\star \sim 1-4\Rsun$, a Neptune- or Jupiter-like planet) are similar to the system we have simulated.


\section{Summary and conclusions} \label{sec:conclusion}
We performed a 3D hydrodynamical simulation of \ac{HJ} engulfment by a 1\Msun star located on the lower part of the red giant branch, where it has expanded to a radius of 4\Rsun. This scenario has been chosen to reflect star-planet interaction at the smallest separation that would still confidently lead to a direct merger, that is, averting planet Roche-lobe overflow or tidal disruption outside the star. It could represent the fate of observed ultra-short period \acp{HJ}, or longer-period Jovian planets that migrate inwards due to tides. In the simulated regime, drag from ram pressure rather than gravitational focussing of upstream gas drives the spiral-in. To self-consistently model ram pressure, and also to capture other effects such as mass ablation and disruption, we modelled the \ac{HJ} as a polytropic gas sphere.

We identify four different phases of engulfment, similar to those outlined by \cite{Metzger+12} and \cite{Jia+Spruit18}. An initial grazing phase is characterised by relatively slow spiral-in near the stellar surface, where the density scale height ranges from a few $\times0.1-1\Rj$. The main engulfment phase proceeds when the \ac{HJ} becomes completely immersed within the stellar envelope, and spirals towards the core in roughly a single orbit. The strong drag force gives rise to a sinking stage where radial in-fall dominates azimuthal motion, ending in disruption by a combination of tides and ram pressure from the dense radiative layers beneath the outer convective zone. The disrupted planetary fragments sink until reaching density equilibrium with their surroundings.

The \ac{HJ} gradually dissipates $\approx 90\%$ of its mass into the convective region, which may enrich the stellar surface with \Li and heavy elements after mixing. Our resolution test suggests that the total amount of mass loss is robust, although the precise amount that is ablated at different separations has not fully converged. In an upcoming paper, we present simulations with converged ablation rates and analyse the ablation mechanism in detail (Lau et al. in prep.). We estimate that \Li is enhanced by 0.13 dex, which is not statistically significant given the intrinsic \ALi variation in open clusters has a similar magnitude. Producing a stronger enrichment signature would require engulfing or accreting more massive substellar bodies (e.g. brown dwarfs) onto stars that have evolved further along the red giant branch, where the baseline \Li abundance and convective mass decrease.

Significant ($\gtrsim 10\kms$) surface rotational velocities are observed at the end of our simulation, although representing a very small amount of mass ($\sim 10^{-3}\Msun$). The bulk of the envelope rotates with $\sim 1\kms$, consistent with the expectation from angular momentum conservation. The maximum induced surface rotational speed scales linearly with the mass of the engulfed object, and so could reach $\sim 10\kms$ from engulfing a more compact, say 10\Mj, \ac{HJ}.

The orbital energy deposited via drag into the stellar envelope may produce a luminous transient, particularly for engulfment events occurring on the subgiant or lower part of the red giant branch. The small amount of unbound ejecta, $\sim10^{-5}\Msun$ in our simulation, could give rise to a second, recombination-powered peak in the lightcurve that is ten times more luminous than the unperturbed star. The latter was recently proposed as the emission mechanism for the transient ZTF SLRN-2020 \citep{De+23} from the engulfment of a Neptune- or Jupiter-mass planet by a main-sequence or slightly evolved Sun-like star. Understanding such transients motivates dedicated studies of ejecta launching and the transport of drag luminosity during the grazing phase and onset of engulfment.



\begin{acknowledgements}
   This work has benefited from useful discussions with Tatsuya Matsumoto, Brian Metzger, Ryosuke Hirai, Andy Casey, and Amanda Karakas. We further thank Melinda Soares-Furtado for sharing the GALAH A(Li) data \citep{Soares-Furtado+21} used in our estimate of the Li enrichment signature. This paper made use of data from the NASA Exoplanet Archive, which is operated by the California Institute of Technology, under contract with the National Aeronautics and Space Administration under the Exoplanet Exploration Program. The simulations presented in this work were performed on the Rusty supercomputer and Popeye supercomputer of the Flatiron Institute, which is supported by Simons Foundation. M. Y. M. L. acknowledges support by an Australian Government Research Training Program (RTP) Scholarship and Monash University Postgraduate Publications Award. I.M. is a recipient of the Australian Research Council Future Fellowship FT190100574. D.P. acknowledges ARC funding via DP220103767 and DP240103290. Parts of this research were supported by the Australian Research Council Centre of Excellence for Gravitational Wave Discovery (OzGrav), through project number CE170100004.

   The \MESA stellar model used in our simulations, the inlist used to generate the model, and data that were used to produce all figures in this article are available at \url{https://dx.doi.org/10.26180/21342090}.
\end{acknowledgements}

%
%
\bibliography{bibliography.bib}{}
\bibliographystyle{aa}

\begin{appendix}
   \onecolumn
\section{Stability of planet model} \label{app:planet-stability}
\begin{figure*}[h!]
    \includegraphics[width=\linewidth]{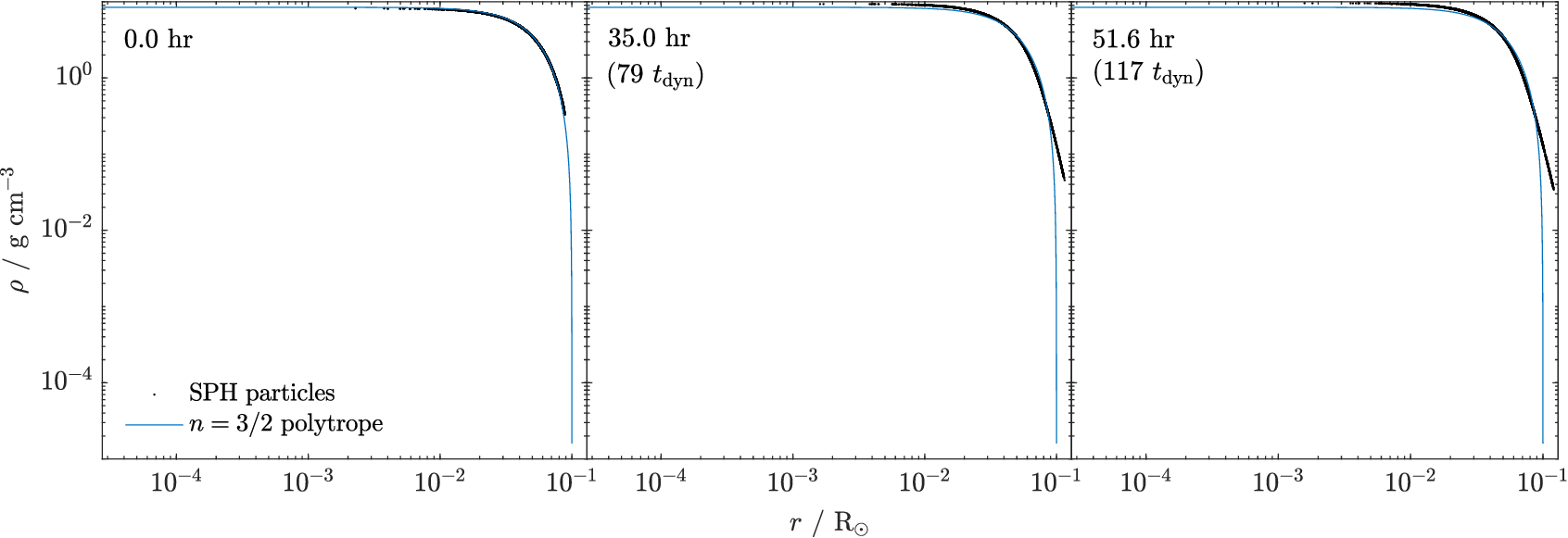}
    \caption{Density distribution of the \ac{HJ} model at three different times when evolved in isolation, showing that the planet preserves its initial structure for a duration close to the planet-disruption time. The times of each profile and the ordinate scale are chosen to match the first three panels of Figure \ref{fig:planet_rho_s_r} to aid comparison. The time in units of the \ac{HJ}'s free-fall time is also displayed in parentheses. Each black dot represents one of $N_\mathrm{p}=12,300$ \SPH particles used to resolve the planet structure. The density profile of an $n=3/2$ polytrope is plotted in blue for comparison.}
    \label{fig:planet-stability}
\end{figure*}

We ensure that the \ac{HJ} model maintains hydrostatic balance and its initial structure to a satisfactory degree by simulating it in isolation up to $t=51.6\hr$, which is near the time of planet disruption in the simulation. This is equivalent to 117 times the \ac{HJ}'s surface free-fall time. Figure \ref{fig:planet-stability} shows the \ac{HJ} density profile at $t=0.0,35.0,$ and $51.6\hr$ compared to the intended $n=3/2$ polytrope (blue line). Due to the inclusion of \SPH thermal conductivity, the planet's surface gradually expands while the centre condenses. The radius and central density increase by a factor of 1.38 and 1.17, respectively, by 51.6\hr.

\FloatBarrier

\section{Resolution study} \label{app:resolution}
To study the resolution dependence of our results, we produce four additional simulations that resolve the star with a different number of \SPH particles: $N_\star = 10^6$, $3\times10^6$, $10^7$, and $3\times10^7$. Due to the use of equal-mass particles in \Phantom, the number of particles that is initially used to resolve the planet in each case is $\Mp N_\star / (\Mstar - M_\mathrm{core}) =$ 1262, 3786, 12616, and 37846, respectively. Because we are most interested in whether the evolution of \Mp and $a$ have converged, we focus on simulating the engulfment phase, where these quantities change the most. We therefore start with the \ac{HJ} being fully immersed in the stellar envelope ($a(t=0)= 3.8\Rsun$) and also use a larger core boundary than in our main simulation, $r_\mathrm{core} = 0.372\Rsun$ and $M_\mathrm{core} = 0.207\Msun$. These measures reduce the dynamic range, making the higher-resolution simulation (with $N_\star= 3\times10^7$) feasible.

\begin{figure*}[h!]
    \centering
    \begin{minipage}[b]{.48\textwidth}
            \includegraphics[width=\linewidth]{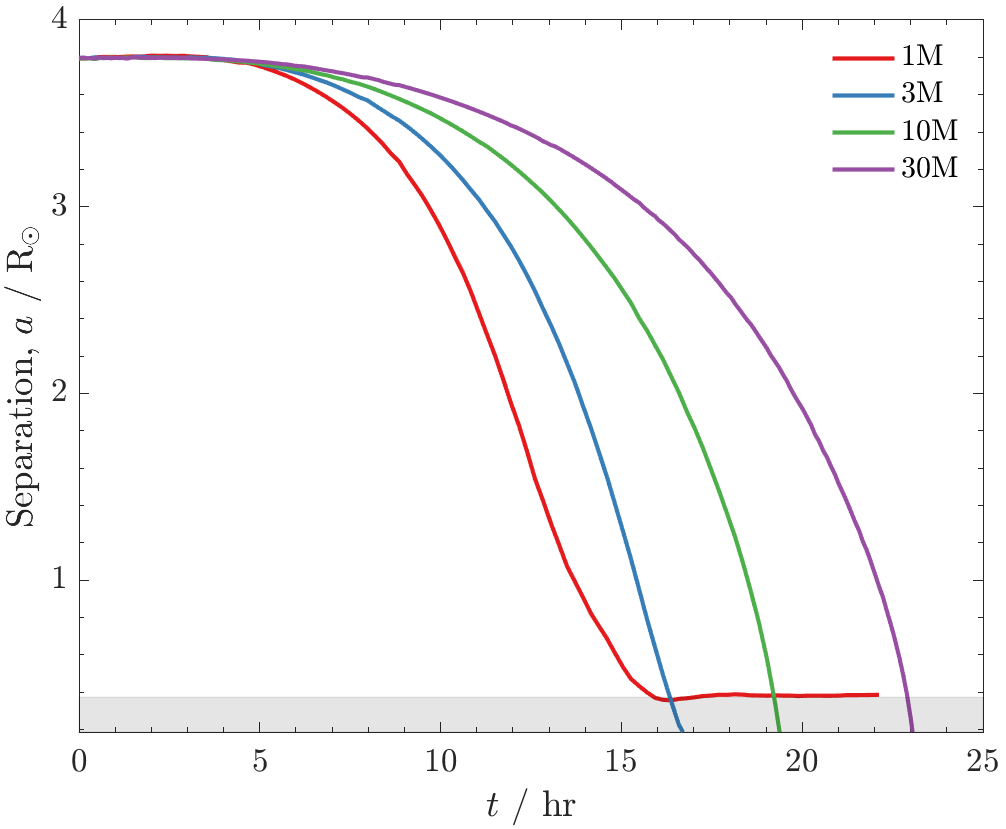}
            \caption{Planet separation, $a$, as a function of time at different resolutions, indicated by the total number of \SPH particles used to resolve the star ($N_\star$) in each simulation. The shaded region is beneath the star's core excision radius.}
            \label{fig:res_sep}
    \end{minipage}\hfill
    \begin{minipage}[b]{.48\textwidth}
            \includegraphics[width=\linewidth]{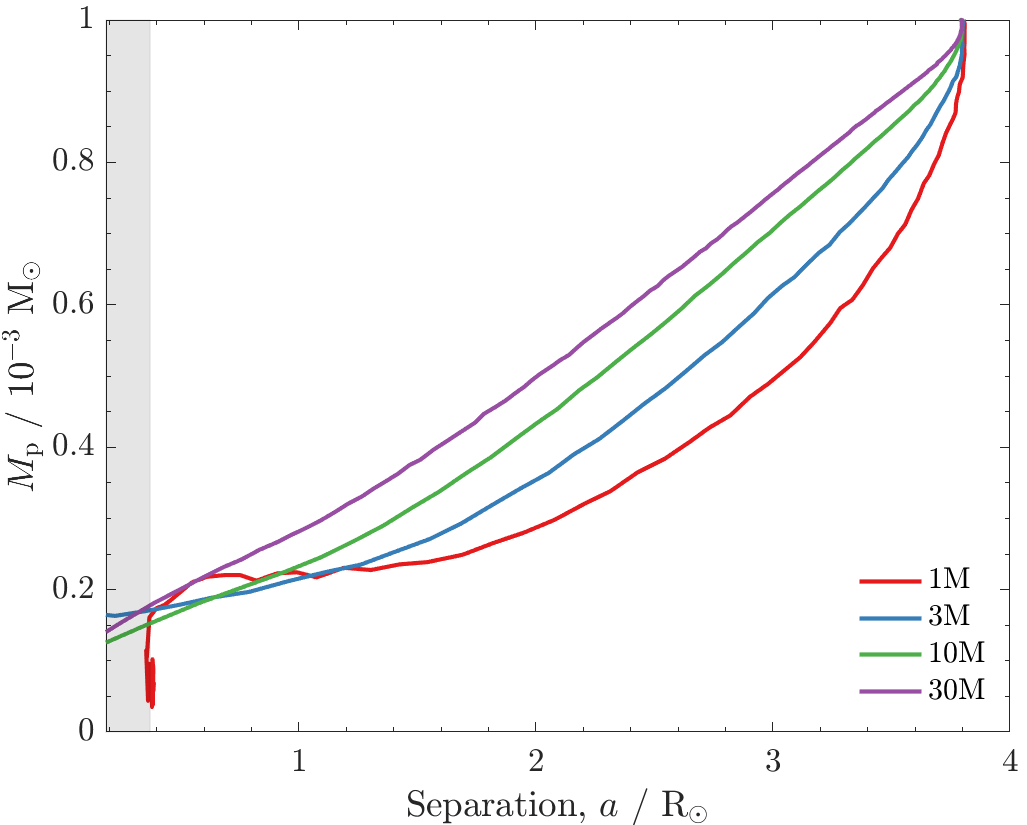}
            \caption{Planet mass, \Mp, as a function of separation, $a$, at different resolutions indicated by the total number of \SPH particles used to resolve the star ($N_\star$) in each simulation. The shaded region is beneath the star's core excision radius.}
            \label{fig:res_mass}
    \end{minipage}
\end{figure*}

Figure \ref{fig:res_sep} shows the time evolution of $a$ at different resolutions. The simulations have yet to converge on the duration of the engulfment phase, with higher resolution resulting in a longer spiral-in. The cause of this has been raised in Section \ref{subsec:phases}---higher resolution better resolves the low-density stellar surface. When setting up the 3D star, under-resolving the pressure scale height at the surface causes slight expansion during the relaxation phase, resulting in a higher density compared to the \MESA model at the same radius. This increases the drag luminosity near the surface, and shortens the length of the grazing phase and transition to full engulfment. We have also restricted Figure \ref{fig:res_sep} to $a>r_\mathrm{core}/2$, as the inspiral beneath the inner boundary has no physical meaning, and the stellar density near $r_\mathrm{core}$ for our resolution study is generally too low for the disrupted planetary material to settle. The $N_\star = 10^6$ simulation (red curve) is an exception, where the densest material stalls at $a\approx 0.37\Rsun$ because the \ac{HJ}'s central density decreases sufficiently over the engulfment process for it to become neutrally buoyant at that radius.

Figure \ref{fig:res_mass} shows the resolution dependence of \ac{HJ} ablation. As discussed in Section \ref{subsec:ablation}, the initial mass loss, concentrated at $a=3.8\Rsun$, is overestimated at low resolution. This could be because the \SPH conductivity that is active when resolving steep density gradients in low-density regions artificially heats the \ac{HJ}. But the ablation rate in the later parts of engulfment decreases to give nearly the same overall mass that is lost to the convective envelope across different resolutions. The \SPH conductivity, which is second order in the particle smoothing length, decreases towards the denser stellar interior. With the $N_\star = 10^7$ and $3\times10^7$ simulations (green and purple curves), the apparent mass loss at the onset of engulfment is no longer present, and \Mp instead decreases linearly with $a$. The final \ac{HJ} mass, near $a=r_\mathrm{core}$, appears robust as it falls within the range of 1.5--2.0$\times 10^{-4}\Msun$ across all tested resolutions.

The amount of unbound mass, which is relevant for understanding the recombination transient discussed in Section \ref{subsec:transients}, changes by a factor of a few across the different simulations. For $N_\star = 10^6$, $3\times10^6$, $10^7$, and $3\times10^7$, the amount of unbound material are 2.0, 1.6, 1.1, and 0.82, respectively, in units of $10^{-5}\Msun$. The amount of unbound ejecta therefore decreases with resolution, likely to be also related to the ability to resolve the low-density stellar surface, where much of the unbound material originates. However, increasing the starting orbital separation, which lengthens the grazing phase, also leads to a larger amount of unbound mass. The default simulation, carried out with $N_\star = 10^7$ and starting at $a(t=0)=4.1\Rsun$, ejects $1.6\times10^{-5}\Msun$ of material, compared to the $1.1\times10^{-5}\Msun$ of material ejected in the resolution test with the same $N_\star$ but starting at $a(t=0)=3.8\Rsun$. The amount of unbound ejecta is therefore a considerable source of uncertainty in our estimate of the plateau luminosity and duration of the H-recombination transient in Section \ref{subsec:transients}.

Overall, the resolution study shows that our simulation likely underestimates the length of the grazing phase and delays the onset of the engulfment stage. However, the conclusion that the planet loses $\approx90\%$ of its mass in the convective envelope is robust, and the mass ablation rate appears to be a constant function of separation. The amount of unbound ejecta decreases with resolution but increases with the initial planet separation, though all lying around $\sim 10^{-5}\Msun$ across the tested resolutions.

\end{appendix}

\end{document}